\documentclass[10pt,conference]{IEEEtran}

\PassOptionsToPackage{
    bookmarksnumbered,
    unicode,
    colorlinks=true,
    linkcolor=blue,
    urlcolor=blue,
    citecolor=blue,
    anchorcolor=blue
}{hyperref}

\PassOptionsToPackage{numbers}{natbib}

\usepackage{changepage}
\usepackage{graphicx} % Required for inserting images

\usepackage{amsthm}
\usepackage{amssymb}
\usepackage{amsmath}
\usepackage{mathtools}
\usepackage{changepage}
\usepackage{url}
\usepackage{quantikz}
\usepackage{braket}
\usepackage{listings}
\usepackage{xcolor}
\usepackage{pythonhighlight}
\usepackage[font=small,skip=0pt]{caption}
\usepackage{subcaption}
\usepackage{pdflscape}
\usepackage{tabularx}
\usepackage{algorithm}
\usepackage{algpseudocode}
\usepackage[inline]{enumitem}
\usepackage[mode=buildnew]{standalone}
\usepackage{orcidlink}
\usepackage{subcaption}
\begin{document}
\title{Sdim: A Qudit Stabilizer Simulator}

% \author{Adeeb Kabir}
% \authornote{Authors contributed equally to this research.}
% \email{ask171@rutgers.edu}
% % \orcid{1234-5678-9012}
% \author{Steven Nguyen}
% \authornotemark[1]
% \email{snn28@scarletmail.rutgers.edu}
% \author{Tijil Kiran}
% \email{tk664@scarletmail.rutgers.edu}
% % \author{Rut Mehta}
% % \email{rm1559@scarletmail.rutgers.edu}
% \author{Yipeng Huang}
% \email{yipeng.huang@rutgers.edu}
% \orcid{0000-0003-3171-6901}
% \affiliation{%
%   \institution{Rutgers University}
%   \city{Piscataway}
%   \state{New Jersey}
%   \country{USA}
% }

% \author{Sohan Ghosh}
% \email{sohghosh@ucdavis.edu}
% \author{Isaac H. Kim}
% \email{ikekim@ucdavis.edu}
% \affiliation{%
%   \institution{University of California, Davis}
%   \city{Davis}
%   \state{California}
%   \country{USA}
% }

% \author{James Keppens\,\orcidlink{0000-0001-5698-9549}}
% \email{james.keppens@imec.be}
% \affiliation{Imec, Leuven, Belgium}
% \affiliation{Department of Electrical Engineering, KU Leuven, Leuven, Belgium}

\author{
    \IEEEauthorblockN{Adeeb Kabir\orcidlink{0009-0001-1287-5512},  Yipeng Huang\orcidlink{0000-0003-3171-6901}\\ Steven Nguyen\orcidlink{0009-0009-2786-6705}, Tijil Kiran\orcidlink{0009-0009-3235-1378}}
    \IEEEauthorblockA{Department of Computer Science \\ Rutgers University \\ \{ask171, yipeng.huang\}@rutgers.edu \\ \{snn28, tk664\}@scarletmail.rutgers.edu}
    \and
    \IEEEauthorblockN{Sohan Ghosh\IEEEauthorrefmark{2}\orcidlink{0000-0003-1318-5502}, Isaac H. Kim\IEEEauthorrefmark{3}\orcidlink{0000-0001-7689-3157}}
    \IEEEauthorblockA{Department of Physics\IEEEauthorrefmark{2} \\Department of Computer Science\IEEEauthorrefmark{3} \\ University of California, Davis\\ \{sohghosh, ikekim\}@ucdavis.edu}
    \and
    \IEEEauthorblockN{James Keppens\IEEEauthorrefmark{4}\orcidlink{0000-0001-5698-9549}, Bart Sor\'ee\IEEEauthorrefmark{5}\orcidlink{0000-0002-4157-1956}}
    \IEEEauthorblockA{Imec \IEEEauthorrefmark{4}\IEEEauthorrefmark{5}\\ Department of Electrical Engineering, \\ KU Leuven\IEEEauthorrefmark{4}\IEEEauthorrefmark{5} \\Department of Physics, \\Universiteit Antwerpen\IEEEauthorrefmark{5} \\ \{james.keppens, bart.soree\}@imec.be}
}

\maketitle % should come after the abstract

\begin{abstract}
Quantum computers have steadily improved over the last decade, but developing fault-tolerant quantum computing (FTQC) techniques, required for useful, universal computation remains an ongoing effort.  
Key elements of FTQC such as error-correcting codes and decoding are supported by a rich bed of stabilizer simulation software such as Stim and CHP, which are essential for numerically characterizing these protocols at realistic scales.  
Recently, experimental groups have built nascent high-dimensional quantum hardware, known as qudits, which have a myriad of attractive properties for algorithms and FTQC.  Despite this, there are no widely available qudit stabilizer simulators.  
We introduce the first open-source realization of such a simulator for all dimensions.  We demonstrate its correctness against existing state vector simulations and benchmark its performance in evaluating and sampling quantum circuits.  
Furthermore, we establish its usefulness in a series of FTQC simulations.  The first is an extension of recent logical error rate and threshold experiments with circuit-level noise, and the second is an error detection protocol meant to support the design of nascent qudit hardware.
This simulator is the essential computational infrastructure to explore novel qudit error correction as earlier stabilizer simulators have been for qubits.
\end{abstract}

% add the paper content here

\section{Introduction}

Despite the remarkable progress in the construction of noisy intermediate-scale quantum (NISQ) systems and the identification of NISQ applications~\cite{Preskill2018quantumcomputingin,NAP25196,google_supremacy,AssessingQClandscape,Realistically_Achieving_Quantum_Advantage}, the future of computing is the construction of fault-tolerant quantum computing (FTQC) systems~\cite{Campbell_nature_review_2017,cacm_error_architecture,cacm_efficient_ftqc}.
The quantum error correction code (QECC), which produces a noise-resistant logical qubit out of many noisy physical qubits, is an object of central study in the FTQC paradigm~\cite{Nielsen_Chuang_2010,Devitt_2013,gottesmanqeccbook}.

The conventional wisdom is that QECCs in FTQC systems will be focused narrowly on stabilizer, topological, surface, and toric codes in qubit quantum systems~\cite{Topological_quantum_memory,surface_code_communication,AutoBraid,surface_synthesis}.
The two-dimensional topology of surface codes is well-suited to the two-dimensional locally connected topologies in prototype superconducting quantum computers.
These codes natively support a subset of operations termed Clifford operators, and any non-Clifford operations are introduced via magic state distillation and injection~\cite{Magic-state_functional_units}.
However, recent work suggests that at least 100 physical qubits are needed to realize a single logical qubit~\cite{googleunderthresh}-and maintaining error rates below threshold at that scale is difficult.
The need for the magic-state distillation protocol adds additional overheads to any FTQC algorithm execution.

\begin{figure}[h]
    \centering
    \includegraphics[width=\linewidth]{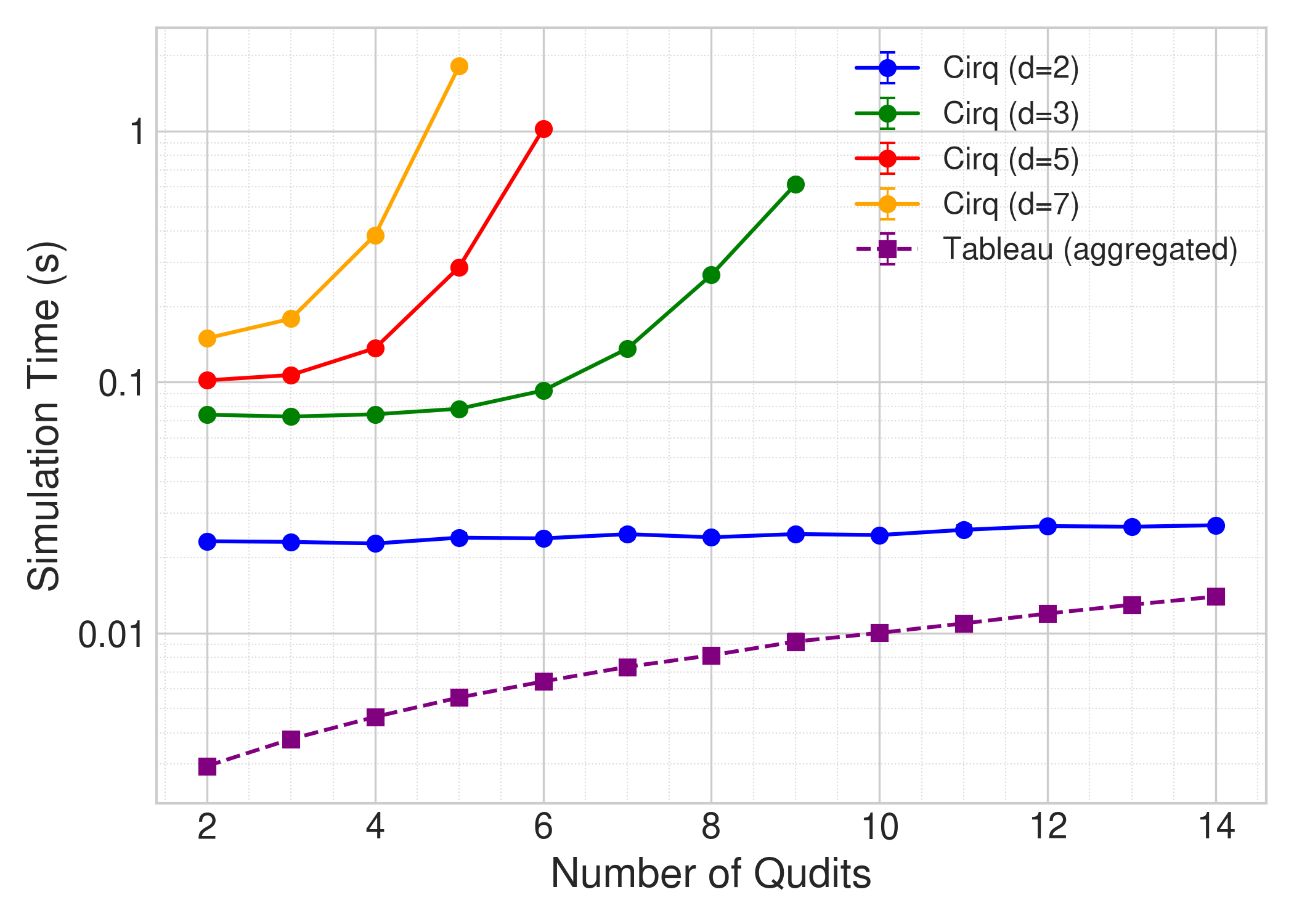}
    \caption{Average time to simulate one shot of a random Clifford circuit using our stabilizer versus Google Cirq statevector simulation.
    The space and time cost of state vector simulation grows exponentially versus the dimension and number of qudits.  In contrast, tableau simulation does not depend on the dimension at all and is bounded by $O(n^2)$, where $n$ is the system size.}
    \label{fig:init-measurement}
\end{figure}

Two trends in quantum device engineering motivate the consideration of novel QECC protocols.
First, emerging quantum devices allow for more dense connections between qubits and further allow for grouped movement of multiple qubits at once.
These abilities enable QECCs where qubit connectivity is less local and is not restricted to a two-dimensional plane~\cite{reconfigurable_atom_arrays}.
Second, most quantum devices support quantum states beyond the lowest two typically used to encode qubits.
These states, termed \textit{qudits}, were once considered much more unreliable to access and control than qubits.
Recent developments have shown remarkable progress in controlling these states~\cite{2020_superconducting_qutrit_hadamard,Goss_superconducting_qutrit_entangling,2022_Seifert_qudit_pulse,2023_two_qutrit_gate,Litteken_2023,2024_Emulating_qubits_ququarts}, and these qudit states have already had an impact on NISQ systems~\cite{PhysRevA.67.012311,PhysRevLett.96.090501,PhysRevA.75.022313,three-level_Toffoli,10.1145/3307650.3322253,9345604,10.1145/3406309,9797403,Compilation_Multi-Level,Litteken_2023,Weggemans2022solvingcorrelation,2023_qudit_qaoa,2024_Emulating_qubits_ququarts}.
The question is what their impact will be on FTQC systems.

While all QECC research begins with the analytical derivation of codes, the computational implementation of the codes is vital to check their correctness and to numerically characterize critical properties such as their physical noise threshold.
Simulations based on the restricted set of Pauli-stabilized quantum states and Clifford gate operations~\cite{gottesman1998heisenbergrepresentationquantumcomputers,Aaronson_2004,Gidney_2021} have been critical for qubit QECC research because, at the scale of physical qubits involved in QECCs, conventional numerical simulation based on matrix vector multiplication on state vectors is prohibitively expensive, seen in Figure \ref{fig:init-measurement}.

On the other hand, such work is absent in qudit QECC research as all available software support for qudit simulation exists only in the non-scalable state vector style~\cite{QuDiet,farias2024quforgelibraryquditssimulation,mato2024mqtquditssoftwareframework,kushnarev2025quditcirqlibraryextension}, despite the theory of qudit stabilizer computation having long since been well understood~\cite{gottesmanhighdim, de_Beaudrap_2013_composite}.

This work introduces an \textit{open-source qudit stabilizer simulator} as the missing piece towards the study of qudit QECCs.  This simulator is \textit{tableau}-based, where the state of $n$ qudits of dimension $d$ is compactly represented in a table of $O(n^2)$ integers compared to the $O(d^n)$ floating point numbers needed for state vectors.  The simulation supports operations for evaluating gates and performing measurements of Pauli operators.

Furthermore, this work contributes an efficient Monte Carlo sampler for measurements, both with and without noise trajectories.  This is achieved via qudit \textit{Pauli frames}, first used in \cite{Gidney_2021} to efficiently sample qubit measurements.  Altogether, this simulator is the minimum needed to efficiently compute large-scale, practical qudit stabilizer error correction circuits on classical hardware.  This efficiency gain traverses the barrier between intractable and tractable, and we will demonstrate this fact by reproducing a code capacity and threshold simulation of a certain qudit error correction code \cite{keppens2025quditvsqubitsimulated} performed in (and limited by) Google Cirq \cite{cirq} in the tableau.  Particularly, we will ``complete'' the original experiment's calculations for $d=7$ where statevector simulation struggles in systems of fewer than 10 qudits.

The simulator was also built concurrently with efforts to design a near-term experiment on a superconducting qutrit QPU, where the goal is to study a five-qutrit patch implementing a small error-detecting code.  Although the target hardware is not large enough to support full active correction, it can support the encoded Clifford gates, stabilizer measurements, and postselected
logical measurements needed to test whether a small qudit code exhibits meaningful logical behavior.  This experimental motivation leads to the LRB-D case study below, where the simulator is used both to validate the circuit construction and to compare the encoded logical qutrit against an unencoded
physical qutrit.

% Initial runs of the protocol in Cirq led to confusing results indicating the code fails at modest depths, and they took nearly a week to sample a paltry number of shots ($\sim10000$) using \textit{only half} of the codespace stabilizer measurements.  However, our simulator took merely \textit{half an hour} to perform the calculation with all stabilizer measurements.   

Finally, we will restrict our attention to prime dimensions $d$.  The simulator supports circuit evaluation for all dimensions $d \geq 2$, but discussion of the non-prime dimensions is technical, is memory inefficient in practice, and is beyond the scope of demonstrating basic utility of qudit stabilizer simulation.  We will address this in future work.  In short, we offer the following contributions in this paper:
\begin{itemize}
    \item We built the first openly available qudit stabilizer simulator functional for all $d \geq 2$.  We will describe its components, operations, and API for any prime dimension $d$.
    \item We will discuss its theoretical correctness and empirically validation against Google Cirq, the most prominent high-dimensional qudit state vector simulator.  Afterward, we will demonstrate that it far out-performs Cirq in computing Clifford circuits in every dimension.
    \item We built a Pauli frame sampler for qudits, which is a protocol that efficiently samples qudit stabilizer circuits.  Its correctness is validated and its performance is benchmarked against Cirq as circuit evaluation was done above.
    \item We will demonstrate the tableau's immediate utility by replicating and extending a recent threshold simulation for a $d = 7$ code under circuit-level noise.  The replication performs significantly faster than statevector, and the circuit-level simulation of high-dimensional quantum codes is the first of its kind.
    \item We will validate a Cirq-simulated error characterization 5-qutrit surface code for upcoming hardware.  This results of the simulation, while prohibitively expensive for Cirq, were matched by the tableau in a significantly shorter period of time.
\end{itemize}

\section{Motivation for characterizing qudit QECC}
% Motivation for computational validation and numerical characterization of qudit QECC

The motivation for computational validation and numerical characterization of qudit QECCs is two-fold.

First, recent experiments in quantum devices are changing the assumptions that have motivated the dominant form of QECCs so far.
The majority of research efforts on the QECCs underlying FTQC systems have been focused on qubit stabilizer surface codes that support Clifford operations within the code~\cite{Topological_quantum_memory}, while non-Clifford operations must be injected through a separate protocol.
However, the steady development of accessing and controlling qudit states~\cite{Wang_qudit,PhysRevA.67.012311,PhysRevLett.96.090501,PhysRevA.75.022313,2008_Manipulating_Biphotonic_Qutrits,2022_qudit_photonic_processor,three-level_Toffoli,Blok2021,Ringbauer_trapped_ion_qudit,Ringbauer_trapped_ion_entanglement,2020_superconducting_qutrit_hadamard,Goss_superconducting_qutrit_entangling,2022_Seifert_qudit_pulse,2023_two_qutrit_gate,Litteken_2023,2024_Emulating_qubits_ququarts} raises the possibility that the first demonstration of FTQC might take place in a qudit system at relatively high physical error rates, rather than in a qubit system at necessarily lower physical error rates~\cite{Campbell_d_level_2014}.

Second, the types of QECC that researchers can fully study (analytically derive, computationally validate, numerically characterize) for now are limited by what classical computers can simulate~\cite{Gidney_2021,keppens2025quditvsqubitsimulated}.
As a result, nearly all QECCs for which there are numerical physical error threshold studies belong to qubit stabilizer codes.
Qudit simulation has received attention from industrial and academic research in the past~\cite{QuDiet,farias2024quforgelibraryquditssimulation,mato2024mqtquditssoftwareframework,kushnarev2025quditcirqlibraryextension}, but all such simulators are based on state vector simulation, and not the scalable variety for stabilizer circuits.

Conventional wisdom posits that qudit codes exhibit higher rates than qubit codes and that they can generally correct more errors \cite{gottesmanqeccbook}, but the lack of efficient simulators have left these claims largely uncontested until recently \cite{keppens2025quditvsqubitsimulated}, and these inquiries are limited by the intractability of statevector simulation.  
Given that stabilizer simulation is what is scalable, it is possible to numerically characterize qudit stabilizer codes.
This paper is the first open-source implementation of such a tool.
\section{Background on qudit stabilizers}

This section, reviews the essentials of quantum stabilizer computation, particularly the representation of quantum states as a table of integers known as a \textit{tableau}.  As an aside, this formalism is exactly the same as describing a stabilizer code~\cite{Gottesman_1998, gottesmanqeccbook}.

\subsection{Qudit Pauli stabilizer states \& Clifford operations}

\paragraph{States} 
A dimension $d$ qudit is a quantum system with $d$ computational basis states $\{ \ket{0}, \dots, \ket{d - 1}\}$.
We write product states in various ways: $\ket{k_1 k_2 \dots k_n} = \ket{k_1} \otimes \ket{k_2} \otimes \dots \otimes \ket{k_n} = \ket{k_1}\ket{k_2} \dots \ket{k_n}$.
All qudit states are either product states or sums of product states.

\paragraph{Gates}
A single qudit gate $G$ acting locally on the qudit $l$ is written $G_l$, and when gates $G_1, \dots, G_m$ act locally on distinct qudits in state $\ket{k_1 k_2 \dots k_n}$, we similarly omit the product symbol for brevity: $G_1 G_2 \dots G_m$.
If $m < n$, then at least one $G_l$ acts on the $q > 1$ qudits, which, as written, means that it acts locally on the product state $\ket{k_{l} k_{l + 1} \dots k_{l + q}}$.

\paragraph{Pauli operators}
The \textit{Pauli group} $\mathcal{P}$ is ubiquitous in quantum computing as both a set of elementary operations and a discrete basis for arbitrary errors~\cite{preskill1998lecture,mermin2007quantum,Nielsen_Chuang_2010,dewolf2023quantumcomputinglecturenotes}.
The qudit Pauli group $\mathcal{P}(d)$ is generated by gates $X$ and $Z$ (identified as \textit{flip} and \textit{phase} errors) where $X \ket{j} = \ket{(j + 1) \mod d}$ and $Z \ket{j} = \omega^j \ket{j}$, where the root of unity $\omega = \exp{(2 \pi i / d)}$.
We note the anti-commutation relation $ZX = \omega XZ$.
All Pauli operators, including the ``$Y$''-type operators can simply be written as a product of $X$ and $Z$ up to phase, so we always express and identify the elements of $\mathcal{P}$ in that form.  
Explicitly, we may write \textit{any} of them as $\omega^{c} X^a Z^b$, identified by a 3-tuple integer encoding (mod $d$) $(a, b, c)$.  Tensor products of Pauli operators are commonly called \textit{Pauli strings}.

\paragraph{Clifford operators}
The qudit \textit{Clifford group} $\mathcal{C} := \mathcal{C}(d)$ consists of all gates $C$ that \textit{normalize} the Pauli group, meaning $CWC^\dagger \in \mathcal{P}$, for any $W \in \mathcal{P}$.
The qubit Cliffords are commonly written as gates generated by $\{ CNOT, H, P\}$ \cite{gottesman1998heisenbergrepresentationquantumcomputers,Aaronson_2004}, each of which is typically generalized as $\{ SUM, \mathcal{F}, P'\}$ for $d > 2$.
These gates act as follows: 
\begin{align*}
    SUM \left( \ket{i} \otimes \ket{j} \right) &= \ket{i} \ket{(i + j) \mod d}
    \\
    \mathcal{F} \ket{i} &= \sum_{j = 0}^{d - 1} \omega^{ij} \ket{j}
    \\
    P'\ket{j} &= \omega^{j (j - 1) / 2} \ket{j}
\end{align*}

The $SUM$ gate conditionally applies $X^i$ on the target wire, and $\mathcal{F}$ is the $d$-dimensional discrete Fourier transform.
The phase gate $P'$ appends a phase that is quadratic in its computational basis label~\cite{gottesmanqeccbook}.  

As an example, consider the state

\begin{equation} \label{stat-state-ex}
    \ket{\phi} = \frac{\ket{00} + \ket{11} + \ket{22}}{\sqrt{3}}
\end{equation}

which is the result of the following Clifford circuit:  

\begin{center}
    \begin{quantikz}
        \ket{0} & \gate{F} & \ctrl{1} &\\
        \ket{0} & \qw & \targ{} &
\end{quantikz}
\end{center}

Clifford operations alone cannot achieve universal quantum computation.
A choice of any single non-Clifford gate, commonly the $T$ gate, operation completes the set of gates needed for universal QC, but simulating this gate set is costly in general \cite{Campbell_nature_review_2017}.

\subsection{Classical simulation of qudit stabilizer circuits}

Clifford operations, while not universal for QC, are expressive enough to describe and simulate the broad class of states that can be protected by (stabilizer) codes.

% \begin{figure}
%     \centering
%     \includestandalone[width=0.9\linewidth]{figures/conj_table}
%     \caption{Conjugation table for Clifford gates on Pauli gates.}
%     \label{fig:conj-table}
% \end{figure}

\paragraph{Stabilizers of a quantum state}
A \textit{stabilizer} $G$ of $\ket{\psi}$ is a Pauli gate $G$ such that $G \ket{\psi} = \ket{\psi}$.
The set of these gates forms a \textit{unique} group $S := S(\ket{\psi})$ under multiplication, called a \textit{stabilizer group}. Returning to example from Equation \ref{stat-state-ex}, we may readily verify that the following operators stabilize $\ket{\phi}$ and are thus elements of $\mathcal{S}(\ket{\phi})$:
\vspace{-2pt}

\begin{equation} \label{stab-state-stabs}
    I_0 I_1, \; X_0^2 X_1^2, \; Z_0^2 Z_1, \; ..., \; X_0^2 Z_0^2  X_1^2 Z_1
\end{equation}

\paragraph{Stabilizer states}
An \textit{$n$-fold stabilizer state} $\ket{\psi}$ is the result of a Clifford circuit applied to $\ket{0}^{\otimes n}$ (also called a \textit{stabilizer circuit}).  Clearly, our favorite example $\ket{\phi}$ is a stabilizer state.  \textbf{A basic but important fact} about $\ket{\psi}$ is that \textit{any} length $n$ list of its non-identity, independent stabilizers $g_1, g_2, \dots, g_n \in S$ \textit{generates} $S$.  For $\ket{\phi}$, every single one of the stabilizers in equation \ref{stab-state-stabs} can be written as products of $X_0^2 X_1^2$ and $Z_0^2 Z_1$.  Furthermore, $X_0^2 X_1^2$ cannot be written as a power of $Z_0^2 Z_1$ and vice versa, so they are independent.  In this way, these two operators generate $\mathcal{S}(\ket{\phi})$.

Another example is $\ket{0}^{\otimes n}$, which is stabilized by operators $Z_0, Z_1, \dots, Z_{n - 1}$.
These are $n$ distinct stabilizers, so their products describe other stabilizers of $\ket{0}^{\otimes n}$ such as $Z_1 Z_2$.  Naturally, this means that the stabilizer generators are a succinct representation of $\ket{\psi}$~\cite{gottesman1998heisenbergrepresentationquantumcomputers}.

% \begin{figure}[h]
%     \centering
%     \includestandalone[width=\linewidth]{figures/conj_table}
%     \caption{The conjugation table that defines rewrite rules when each.  Pauli gates act on the tableau at most by affecting its phase.  While the rules for the \textit{inverse} gates are absent, they are easily derived by reversing the arrows on the gates above.}
%     \label{fig:conjugation-table}
% \end{figure}

% \begin{figure}[hb]
%     \centering
%     \includegraphics[width=0.9\linewidth]{figures/conj_table.png}
%     \caption{Conjugation table for Clifford gates on Pauli gates.}
%     \label{fig:conj-table}
% \end{figure}

% \begin{figure}[ht]
%     \centering
%     \includegraphics[width=0.9\linewidth]{rewriterules.png}
%     \caption{Tableau rewrite rules corresponding to the conjugation table in Figure \ref{fig:conj-table}.}
%     \label{fig:rewrite-rule}
% \end{figure}

\begin{figure}[h]
  \centering
  % center the main caption text
  \captionsetup{justification=centering}

  \begin{subfigure}{\columnwidth}
    \centering
     \includegraphics[width=0.85\linewidth]{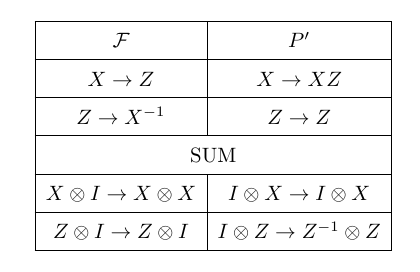}
    \caption{Conjugation table for Clifford gates ($\mathcal{F}, P', SUM$) acting on Pauli gates ($X, Z$).}
    \label{fig:conj-table}
  \end{subfigure}

  \begin{subfigure}{\columnwidth}
    \centering
   \includegraphics[width=0.95\linewidth]{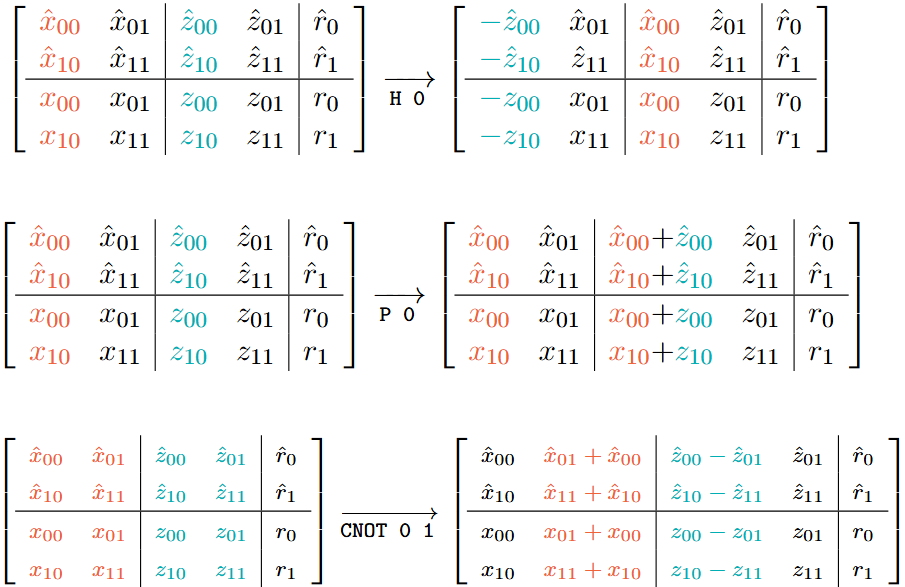}
    \caption{Tableau rewrite rules corresponding to \ref{fig:conj-table}.  Hatted values are for destabilizers while the others are for stabilizers}
    \label{fig:rewrite-rule}
  \end{subfigure}

  \caption{Clifford conjugation of Pauli operators.}
  \label{fig:conj-and-rewrite-rules}
\end{figure}

\paragraph{Stabilizer state evolution}

Now, suppose that $C$ is a Clifford gate and that we want to describe $\ket{\psi} = C \ket{0}^{\otimes n}$ via stabilizer group generators.
Notice:
\vspace{-6pt}

$$\ket{\psi} = C \ket{0}^{\otimes n} = C Z_j \ket{0}^{\otimes n} = (C Z_j C^\dagger) C \ket{0}^{\otimes n} = (C Z_j C^\dagger) \ket{\psi}$$

Operator $C Z_j C^\dagger$ is Pauli by definition, and from the above equation, the operators $\{ C Z_1 C^\dagger, \dots, C Z_n C^\dagger\}$ stabilize $\ket{\psi}$, so they are valid stabilizer generators for $\ket{\psi}$.  \textit{Clifford evolution is a game of conjugating stabilizer generators.}  Figure \ref{fig:conj-table} lists all ways to conjugate Pauli operators.

\paragraph{Stabilizer integer encoding}
Recall that every Pauli $P = \omega^{c} X^a Z^b$.  Every product Pauli $P_1 \otimes \dots \otimes P_m$ can then be written $\omega^{r} (X^{a_1} Z^{b_1} \otimes \dots \otimes X^{a_1} Z^{b_1})$, where $r = c_1 + \dots + c_m$.  We can write it out in a compact \textit{block form} (sometimes known as its \textit{symplectic form} \cite{gottesmanqeccbook}):

\vspace{-6pt}
\begin{align*}
        \left[ \begin{array}{ccc|ccc|c}
        a_0 & \dots & a_0 & b_1 & \dots & b_m & r
        \end{array} \right]
\end{align*}

The two leftmost blocks are the $X$-block and $Z$-\textit{block}, respectively, followed by the \textit{phase block}.  
The product of two stabilizers in the symplectic form is simply the entry-wise sum of their respective symplectic forms.  The conjugation of a stabilizer by a Clifford operator amounts to simple rewrite rules across the blocks, which are derived from the table in Figure \ref{fig:conj-table} into the form in Figure \ref{fig:rewrite-rule}.

%  Finally, the \textit{symplectic inner product} is an important quantity relating two Pauli strings $P_1, P_2$ in block forms:

%  \begin{gather*}
%      P_1 = \left[ \begin{array}{cc|cc|c} a_0 & \dots &  \dots & b_{n - 1}  & r_1\end{array} \right] \\
%      P_2 = \left[ \begin{array}{cc|cc|c} \alpha_0 & \dots  & \dots & \beta_{n - 1} & r_2 \end{array} \right] \\
%      \left< P_1, P_2 \right> = \sum_{k = 0}^{n - 1} b_k \alpha_k - a_k \beta_k  \;\;\; (\text{mod } d)
%  \end{gather*}
 
% This product is 0 if and only if the two strings commute.

\definecolor{codegreen}{rgb}{0,0.6,0}
\definecolor{codegray}{rgb}{0.5,0.5,0.5}
\definecolor{codepurple}{rgb}{0.58,0,0.82}
\definecolor{backcolour}{rgb}{0.95,0.95,0.92}

\lstdefinestyle{python}{
    backgroundcolor=\color{backcolour},   
    commentstyle=\color{codegreen},
    keywordstyle=\color{magenta},
    numberstyle=\tiny\color{codegray},
    stringstyle=\color{codepurple},
    basicstyle=\ttfamily\footnotesize,
    breakatwhitespace=false,         
    breaklines=true,                 
    captionpos=b,                    
    keepspaces=true,                 
    numbers=left,                    
    numbersep=5pt,                  
    showspaces=false,                
    showstringspaces=false,
    showtabs=false,                  
    tabsize=2
}

\lstset{style=python}

% \begin{figure}[t]
%     \centering
% \begin{lstlisting}[language=Python]
% is_constant = random.choice(["True", "False"])
% c = Circuit(2, d)  # Initial circuit is |00>
% c.add_gate("H", 0) # F |0> on qudit 0
% c.add_gate("X", 1) # Flip ancilla |0> to |1> 
% c.add_gate("H", 1) # F |1> on ancilla
% if is_constant:
%     function_constant = random.choice([j for j in range(d)])
%     for i in range(function_constant): 
%         c.add_gate("X", 1) # Oracle for f=j
% else:
%     c.add_gate("CNOT", 0, 1) # Oracle f=id
% c.add_gate("H_INV", 0) # F^t |0> on qudit 0
% c.add_gate("M", 0) # Measure qudit 0
% expected = 0 if is_constant else d - 1
% p = Program(c)
% assert p.simulate() == [MeasurementResult(0, True, expected)]
% \end{lstlisting}
%     \caption{A Sdim program to run and validate the $d$-dimensional Deutsch-Jozsa algorithm.}
%     \label{fig:deutsch_program}
% \end{figure}

\begin{figure}[t]
  \centering
  % center the main caption text
  \captionsetup{justification=centering}

  \begin{subfigure}{\columnwidth}
    \centering
     \begin{lstlisting}[language=Python]
is_constant = random.choice(["True", "False"])
c = Circuit(2, d)  # Initial circuit is |00>
c.add_gate("H", 0) # F |0> on qudit 0
c.add_gate("X", 1) # Flip ancilla |0> to |1> 
c.add_gate("H", 1) # F |1> on ancilla
if is_constant:
    function_constant = random.choice([j for j in range(d)])
    for i in range(function_constant): 
        c.add_gate("X", 1) # Oracle for f=j
else:
    c.add_gate("CNOT", 0, 1) # Oracle f=id
c.add_gate("H_INV", 0) # F^t |0> on qudit 0
c.add_gate("M", 0) # Measure qudit 0
expected = 0 if is_constant else d - 1
p = Program(c)
assert p.simulate() == [MeasurementResult(0, True, expected)]
\end{lstlisting}
    \caption{A program to run and validate the $d$-dimensional Deutsch-Jozsa algorithm on our tableau simulator.}
    \label{fig:deutsch_program}
  \end{subfigure}

  \begin{subfigure}{\columnwidth}
    \centering
   \begin{quantikz}
        \ket{0} & \gate{H} & & \gate[2]{U} & \gate{H^\dagger} & \meter{} \\ 
        % \ket{0} & \gate{\mathcal{F}} &  & & \gate{\mathcal{F}^\dagger} &  \meter{} \\ 
        % \vdots \\
        \ket{0} & \gate{X} & \gate{H} & & & 
    \end{quantikz}
    \caption{Quantum circuit corresponding to the code in Figure \ref{fig:deutsch_program}, which implements Algorithm \ref{alg:qudit-deutsch-algo}.}
    \label{fig:deutsch-circuit}
  \end{subfigure}

  \caption{Tableau program and circuit for the (restricted) Deutsch-Jozsa problem.}
  \label{fig:deutsch-program-and-circuit}
\end{figure}

\section{Validation and evaluation of qudit stabilizer tableau simulation}
% \yipeng{This is the first of two mini-papers in this manuscript. The rhetorical goal of this mini-paper is to teach how qudit stabilizer simulation works, give a corresponding example in Sdim, and demonstrate the performance of stabilizer simulation for qudits.}

The tableau is best understood in action.
Throughout this section, we will introduce its key components and features.  As a guided example, we will demonstrate how these components arise in a quantum circuit, namely a qutrit variant of the \textit{Deutsch-Josza algorithm}.

% \begin{figure}[ht]
%     \centering
%     \includegraphics[width=\linewidth]{deutschcode.pdf}
%     \caption{A sdim program that simulates the Deutsch-Jozsa algorithm and validates its correctness.}
%     \label{fig:deutsch_program}
% \end{figure}

\subsection{Implementation of our open source qudit stabilizer tableau simulator}

Our simulator is a Python library that that creates and evaluates quantum circuits.  The program in Figure \ref{fig:deutsch_program} shows how to write a circuit with most of the major user features and how to read out a measurement.

\paragraph{Data primitives}  
The basic data the user manipulates is a \verb|Circuit| object.  To create an an empty $n$-qudit circuit of dimension $d$, we invoke the constructor \verb|c = Circuit(n, d)|.  The object is initialized to the state $\ket{0}^{\otimes n}$.
Once the user has written out a circuit using the operations detailed below, they can simulate it by creating a \verb|Program| object constructed out of a circuit.  This is seen in line 15 of Figure \ref{fig:deutsch_program}.  

\paragraph{Gate operations and measurements}
The user appends gate instructions to the circuit via its \verb|add_gate| function, which takes arguments for a gate name and a local qudit index (and additional target for \verb|CNOT| when applicable), seen in lines 3-5, 11 of Figure \ref{fig:deutsch_program} .
The elementary gate set is \verb|{X, Z, SUM, F, P}| along with their inverse gates \verb|{X_INV,| \verb|Z_INV, SUM_INV, F_INV, P_INV}|.
The gates can be called by either their qubit or qudit names (e.g. \verb|H| versus \verb|F| on line 3 of Figure \ref{fig:deutsch_program}) with no affect on the program.
At any time, the user can reset the $j^{\text{th}}$ physical qudit to $\ket{0}$ using the \verb|RESET| gate on index $j$.  All measurements at a local index $j$ are $Z$ measurements and are denoted by an \verb|M| instruction.

\paragraph{Error channels}
Our simulator has a gate \verb|N1| that implements common single-qudit noise channels.  Given a positional parameter $j$, a \textit{channel} parameter \verb|t|, and a probability \verb|p|, the code:

\begin{center}
    \verb|c.add_gate('N1', noise_channel=t, prob=p)| 
\end{center}

applies noise channel of type \verb|t| to qudit $j$ with probability \verb|p|, where the types are \verb|'f', 'p', 'd'| corresponding to \textit{flip}, \textit{phase}, and \textit{depolarizing} noise, respectively.

% \begin{figure}[t] 
%     \centering
%     \begin{quantikz}
%         \ket{0} & \gate{H} & & \gate[2]{U} & \gate{H^\dagger} & \meter{} \\ 
%         % \ket{0} & \gate{\mathcal{F}} &  & & \gate{\mathcal{F}^\dagger} &  \meter{} \\ 
%         % \vdots \\
%         \ket{0} & \gate{X} & \gate{H} & & & 
%     \end{quantikz}
%     \caption{Quantum circuit corresponding to the code in Figure \ref{fig:deutsch_program}, which implements Algorithm \ref{alg:qudit-deutsch-algo}.}
%     \label{fig:deutsch-circuit}
% \end{figure}  

\paragraph{Simulation}
The user invokes \verb|simulate| on a \verb|Program| object to evaluate the circuit.  There is an optional \verb|shots| parameter that allows one to sample the circuit many times, which we will return to later.
The simulation produces a 3D list of \verb|MeasurementResult| objects, indexed by shot number, the index of the measurement, and sequential order of the measurement in the circuit.
The \verb|MeasurementResult| itself stores the qudit index, the measurement outcome, and whether it was deterministic.  If only a single shot is simulated, the list is 2D.  Line 16 of Figure \ref{fig:deutsch_program} demonstrates checking whether a single terminal measurement of Deutsch-Jozsa on the first qudit is either 0 or $d - 1$.

\begin{figure*}[ht]
    \centering
    \includegraphics[width=0.80\linewidth]{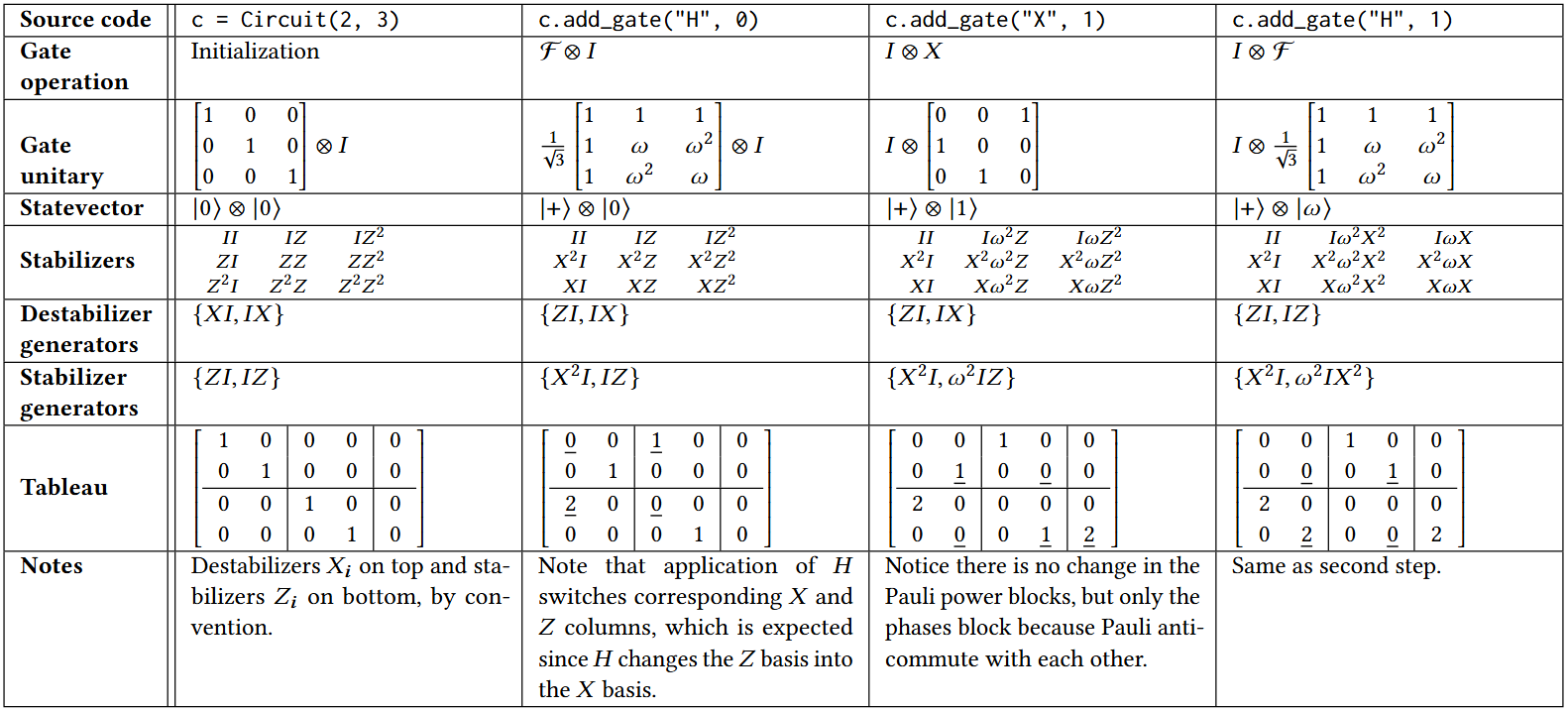}
    \caption{The first three steps of the Deutsch-Jozsa algorithm written out in various representations, including the tableau.}
    \label{table:D-J-3-steps}
\end{figure*}

\subsection{Demonstration of program architecture using qutrit Deutsch-Jozsa}

 Algorithm \ref{deutsch-algo-proc} solves a heavily restricted version of the Deutsch-Jozsa problem, a classic quantum oracle task ~\cite{qudit_deutsch_jozsa,Wang_qudit}, for qutrits where $n = 1$, and the input function $f$ is either constant or the identity function.  Figure \ref{fig:deutsch_program} exactly implements the circuit in Figure \ref{fig:deutsch-circuit} as an tableau program.

\begin{algorithm}
\caption{(Restricted) Qudit Deutsch-Jozsa in dimension $d$}\label{alg:qudit-deutsch-algo}
\begin{algorithmic}[1]
    \Require Function $f$, either constant or identity
    \Ensure 0 if constant, balanced otherwise

    \State Initialize $\ket{0} \ket{1}$ and apply $\mathcal{F} \otimes \mathcal{F}$ to it.
    \If{ function $f$ is constant}
        \State $U_f = X_1^j$, where $f(x) = j$, for every input $x$
    \Else
        \State $U_f = CNOT_{0, 1}$
    \EndIf
    \State Apply $U_f$ and then $\left( \mathcal{F}^\dagger \right)$.
    \State Return the outcome of measuring $Z_0$.
\end{algorithmic} \label{deutsch-algo-proc}
\end{algorithm}

\subsubsection{Operations in the prime dimension qudit stabilizer tableau}

The \textit{tableau} of a $n$-qudit stabilizer state $\ket{\psi}$ is a table recording $n$ of its stabilizer generators stacked in block form.  As discussed in the background, the tableau and its change under unitary gates is an exact representation of the state we are simulating.

In practice, we append additional data to the tableau in the form of the ``\textit{destabilizers}'' of $\ket{\psi}$, or Pauli operators that \textbf{do not} stabilize the state \textit{up to any phase}.  
For example, $\ket{1}$ is not stabilized by $Z$, but it is stabilized by $\omega^{-1} Z$, so $Z$ is not a destabilizer generator.  However, $X$ cannot stabilize $\ket{1}$ even with additional phases, so $X$ is a generator for other destabilizers of $\ket{1}$ (such as $X^2, X^3, \dots$).  Such operators form a group (analogous to $\mathcal{S}(\ket{\psi})$) and can be expressed via $n$ generators. 
Destabilizers were the key in seminal work  in qubit stabilizer simulation to achieve $O(n^2)$ time for measurement ~\cite{Aaronson_2004,gottesmanhighdim}.
In total, the tableau consists of $2n$ Pauli strings in block form.  
The destabilizer block is written on top of the stabilizer block, and so altogether, the tableau takes the following form with $O(n^2)$ integers:
\vspace{-6pt}

\begin{equation} \label{tab-state}
    {\setlength{\arraycolsep}{1.5pt}
    \left[ \begin{array}{ccc|ccc|c}
        x_{0, 0} & \dots & x_{0, n - 1} & z_{0, 0} & \dots & z_{0, n - 1} & r_0 \\
        \vdots & \ddots & \vdots & \vdots & \ddots & \vdots & \vdots \\
        x_{n - 1, 0} & \dots & x_{n - 1, n - 1} & z_{n, 0} & \dots & z_{n - 1, n - 1} & r_{n - 1} \\ \hline
        x_{n , 0} & \dots & x_{n , n} & z_{n, 0} & \dots & z_{n, n - 1} & r_{n} \\
        \vdots & \ddots & \vdots & \vdots & \ddots & \vdots & \vdots \\
        x_{2n - 1, 1} & \dots & x_{2n - 1, n - 1} & z_{2n - 1, 0} & \dots & z_{2n - 1, n - 1} & r_{2n - 1} \\
        \end{array} \right] }
\end{equation}

 These tableaus are built as Numpy \cite{harris2020array} arrays.  We initialize the tableau as $\ket{0}^{\otimes n}$ using the destabilizer and stabilizer generators $\{X_j \}_{j = 1}^{n}$ and $\{Z_j \}_{j = 1}^{n}$, respectively, by convention.  Evolving $\ket{\psi}$ by a gate $C$ conjugates all stabilizer and destabilizer generators by $C$ via its rewrite rules in Figure \ref{fig:rewrite-rule}. 

 The first three steps of the circuit in Figure \ref{fig:deutsch-circuit} are detailed in Table \ref{table:D-J-3-steps}.  Each row explicitly states information from the tableau or an equivalent form in another representation.  Now that we can properly evolve a tableau via gates, we discuss how to extract measurements (and post-measurement states) out of it.

\subsubsection{Measurement in the prime dimension qudit stabilizer tableau}
Our ($Z$) measurement algorithm is detailed in Algorithm \ref{alg:meas}.  It suffices to perform any Pauli measurement; given $P = U Z U^\dagger$, we may perform a general Pauli $P$ measurement by applying  $U^\dagger$, performing a $Z$ measurement, and applying $U$ \cite{gottesmanqeccbook}. Lines 2-8 describe the procedure for a \textit{random} measurement, while lines 10-15 detail a \textit{deterministic} one.  

\begin{algorithm}[ht]
\caption{Measurement of a qudit of (prime) dim. $d$}\label{alg:meas}
\begin{algorithmic}[1]
    \Require Meas. index $j$, $n$-qudit state $\ket{\psi}$ in tableau form
    \Ensure Tableau after measurement on $\ket{\psi}$ with outcome $\alpha$

    \State $\alpha \gets 0$
    \If{$\exists i, j$ such that $i \geq n$ and  $x_{i, j} \neq 0$ and $x_{i, l} = 0$, $\forall l < j$}
        \State \textbf{for} $k > i$ \textbf{such that} $x_{k, j} \neq 0$
        \State \ \ \ \ let $h$ be such that $h \cdot x_{i, j} + x_{k, j} \equiv 0 \; (\text{mod } d)$
        \State \ \ \ \ row $k \gets$ row $(k + h) \;\cdot$ row $i$ 
        \State $\alpha \gets $ random integer (mod $d$).
        \State destab. row $(i - n)$ $\gets$ stab. row $i$.
        \State stab. row $i$ $\gets$ block form of $\omega^{-\alpha} Z_j$
    \Else 
        \State let $\gamma = I^{\otimes n}$
        \State \textbf{for} $i  < n$ \text{such that} $x_{i, j} \neq 0$
        \State \ \ \ \ let $\varphi = \text{Pauli string of stab. row } i + n$
        \State \ \ \ \  $\gamma \gets \gamma \; \cdot \;\varphi^{x_{i, j}}$ 
        \State let $\phi = \text{Phase of } \gamma$
        \State $\alpha \gets - \phi / 2$ 
    \EndIf
    \State \Return $\alpha$ and tableau
\end{algorithmic}
\end{algorithm}

\paragraph{Measurement with random outcomes}
If a $Z_j$ measurement of the tableau for $\ket{\psi}$ is random, then $Z_j \notin \mathcal{S}(\ket{\psi})$.  At least one stabilizer anti-commutes with $Z_j$ (we can rewrite the tableau so exactly one of them does \cite{gottesmanhighdim}), which is true when some $x_{i, j} \neq 0$ in the stabilizer block.  We replace this stabilizer with the block form of $\omega^{-r} Z_j$, where $r$ is random, and the old stabilizer becomes the $j^{\text{th}}$ destabilizer.  By definition, this new tableau is a state that has collapsed to the $\omega^r$-eigenstate of $Z_j$.  Performing this is at most $O(n^2)$ steps if more than one stabilizer anti-commutes.

\paragraph{Measurement with deterministic outcomes}
Otherwise, $cZ_j \in \mathcal{S}(\ket{\psi})$, where $c = \omega^l$.  Algorithm \ref{alg:meas} calls to multiply all stabilizers such that their corresponding destabilizers anti-commute with $Z_j$ and return half of the inverse of the product's overall phase as the outcome.  The destabilizer method for deterministic stabilizers measurements is a straightforward generalization of Aaronson and Gottesman's procedure for qubits \cite{Aaronson_2004, gottesmanhighdim} by noting that the algorithmic invariants remain the same in all prime dimensions.  The outcome does not affect the tableau, which is consistent with the fact that the stabilizer state is an eigenstate of the measurement operator.

% We will briefly prove the procedure in \ref{alg:meas} determines $c$ and thus the measurement outcome.

% Let $S_0(t), \dots, S_{n - 1}(t)$ and $D_0(t),\dots, D_{n - 1}(t)$ denote the stab. and destab. generators of a tableau at time step $t$.  We initialize the tableau to $S_i(0) = Z_i$ and $D_i(0) = X_i$.  Notice that $S_i(0)$ anti-commutes with $D_i(0)$ but commutes with $D_j(0)$ where $i \neq j$.  This remains true after evaluating any number of gates $U = C_t C_{t - 1} \dots C_{1}$ since: 
%     \vspace{-6pt}
    
%     \[
%         S_i(t) D_i(t) - D_i(t)S_i(t) = U \Big( S_i(0) D_i(0) - D_i(0)S_i(0) \Big) U^\dagger
%     \]

%     so we can drop the $t$ variable.  We know $c Z_j \in \mathcal{S}(\ket{\psi})$, meaning $cZ_j = \prod_{k = 0}^{n - 1} \gamma_kS_k$ by definition.  Then we apply the (anti-)commuting conditions from before to calculate:
%     \vspace{-6pt}

%     \begin{gather*}
%         x_{i, j} = \left< D_i, Z_j\right> = \sum_{k = 0}^{n - 1} \gamma_k \left< D_i, S_k\right> = \gamma_i \left< D_i, S_i\right> \\
%         \Rightarrow \gamma_i = x_{i, j} \cdot \big( \left< D_i, S_k\right> \big)^{-1}
%     \end{gather*}

%     Multiplying all the stabilizers and $\omega^{\gamma_i}$ terms together yields a Pauli string with overall phase $c = \omega^l$, where $-l$ is the measurement outcome.  This takes $O(n^2)$ time, and the proof above is a generalization of the one given in \cite{Aaronson_2004} for qubits.

 \begin{figure}
    \centering
    \begin{quantikz}
        \ket{0} & \gate{H} & \ctrl{1} & \qw & \meter{} & \\
        \ket{0} & \qw & \targ{} & \meter{} & \qw & 
    \end{quantikz}
    \caption{A simple circuit with two measurements.  The first is random, and the second is deterministic and matches the first.}
    \label{fig:meas-example}
\end{figure}  

The circuit in Figure \ref{fig:meas-example} following is an example an circuit with a random measurement followed by a deterministic one.  Skipping the first two gates, we write out the tableau right before its terminal measurements

\begin{equation} \label{eqn:meas-example}
        \left[ \begin{array}{cc|cc|c}
0 & 0 & 1 & 0 & 0 \\
0 & 1 & 0 & 0 & 0 \\ \hline
2 & \textcolor{red}{2} & 0 & 0 & 0 \\
0 & 0 & 2 & 1 & 0 
\end{array} \right] 
\xrightarrow[\texttt{M 1}]{}
    \left[ \begin{array}{cc|cc|c}
2 & \textcolor{red}{2} & 0 & 0 & 0 \\
0 & 1 & 0 & 0 & 0 \\ \hline
0 & 0 & 0 & 1 & \textcolor{red}{r} \\
0 & 0 & 2 & 1 & 0 
\end{array} \right] \xrightarrow[\texttt{M 0}]{}
\end{equation}

The first $Z_1$ measurement is of the left tableau in Expression \ref{eqn:meas-example}.  In this case, there is only one non-commuting row, with the offending $x_{i, 1}$ value marked in red.  This row becomes the corresponding \textit{destabilizer} in row $i - n$.  This row is finally replaced by a block form of $\omega^r Z_1$, where $r$ is random.  For the second $Z_0$ measurement, we now destabilizer rows of the rightmost tableau.  In this case, only row $0$ does not commute with $Z_0$, and after performing the calculations, we simply obtain outcome $r$.

%\input{4.3_nonprime}
% \begin{figure}[hb]
%     \centering
%     \includestandalone[width=0.8\linewidth]{figures/b-v-circuit}
%     \caption{A qudit Bernstein-Vazirani circuit where the secret string starts with ``10''.  For every nonzero character in the string, a qudit in that character's position entangled with the bottom ancilla via the CNOT oracle.}
%     \label{fig:b-v-circuit}
% \end{figure}

% \begin{figure}[h]
%     \centering
%     \includegraphics[width=0.9\linewidth]{b-v-times-FINAL.png}
%     \caption{Cirq against the tableau on evaluating (the first shot of) Bernstein-Vazirani circuits.  Each test is randomized over 100 bitstrings with the same amount of 1s, which controls the size of the most expensive subspace to measure.  Note that the time is in logscale, highlight the $O(n^2)$ cost of tableau simulation versus the $O(d^n)$ cost of statevector methods.}
%     \label{fig:b-v-times}
% \end{figure}

% \begin{figure}[hb]
%     \centering
%     \includegraphics[width=0.8\linewidth]{figures/b-v-circuit.png}
%     \caption{A qudit Bernstein-Vazirani circuit where the secret string starts with ``10''.  For every nonzero character in the string, a qudit in that character's position entangled with the bottom ancilla via the CNOT oracle.}
%     \label{fig:b-v-circuit}
% \end{figure}

\begin{figure}[t]
  \centering
  % center the main caption text
  \captionsetup{justification=centering}

  \begin{subfigure}{\columnwidth}
    \centering
    \includegraphics[width=0.9\linewidth]{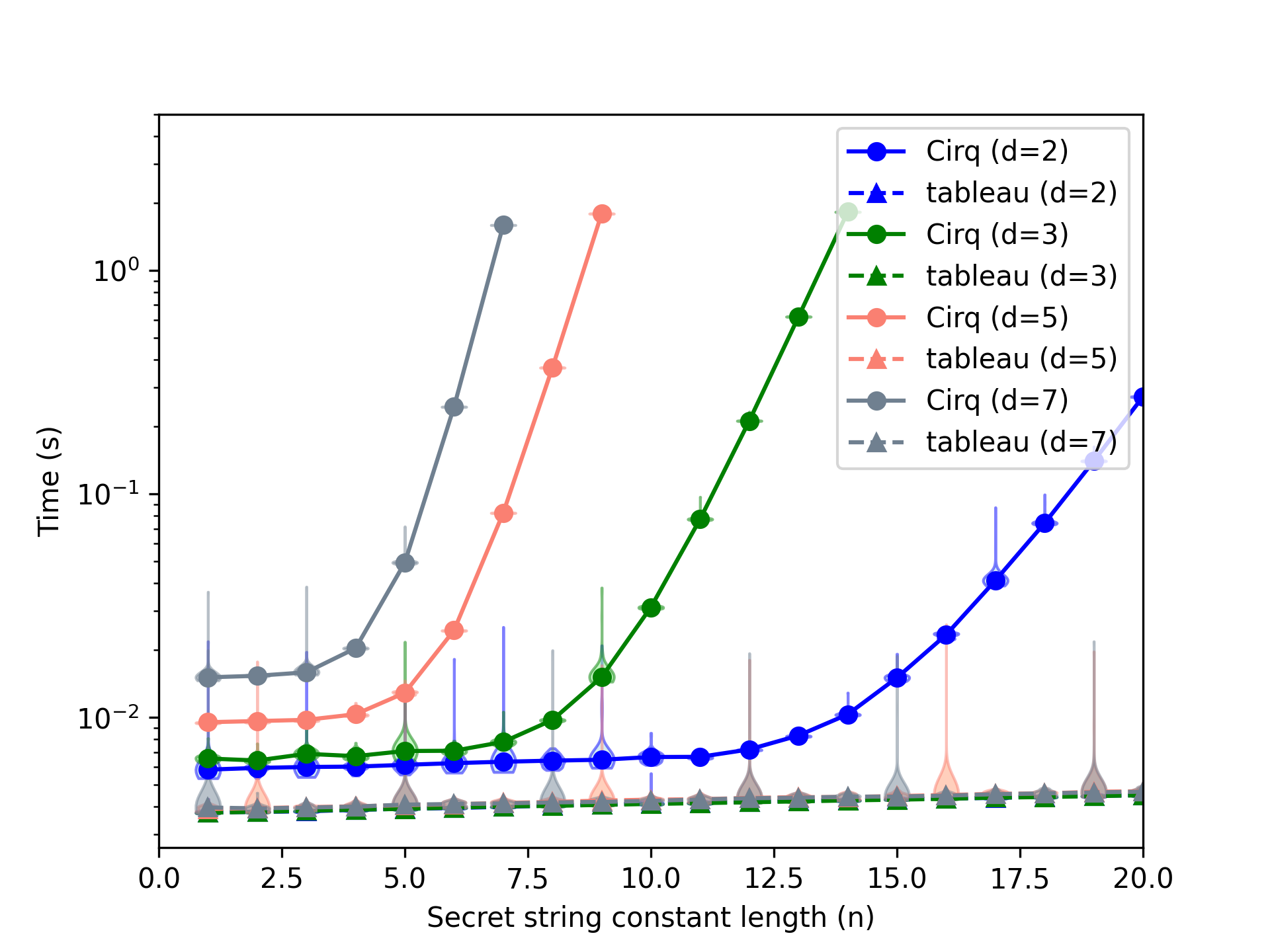}
    \caption{Cirq against the tableau on evaluating (the first shot of) Bernstein-Vazirani circuits.  Each test is randomized over 100 bitstrings with the same amount of 1s, which controls the size of the most expensive subspace to measure.  Note that the time is in logscale, highlight the $O(n^2)$ cost of tableau simulation versus the $O(d^n)$ cost of statevector methods.}
    \label{fig:b-v-times}
  \end{subfigure}

  \begin{subfigure}{\columnwidth}
    \centering
    \includegraphics[width=0.8\linewidth]{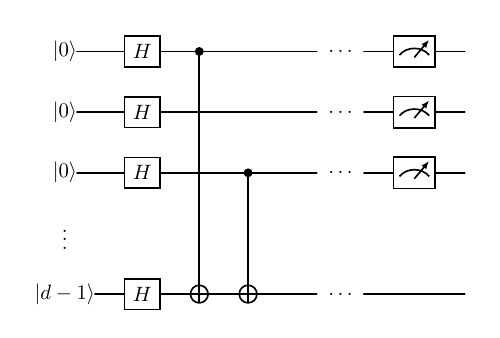}
    \caption{A qudit Bernstein-Vazirani circuit where the secret string starts with ``10''.  For every nonzero character in the string, a qudit in that character's position entangled with the bottom ancilla via the CNOT oracle.}
    \label{fig:b-v-circuit}
  \end{subfigure}

  \caption{Testing the time to initial measurement using Bernstein-Vazirani circuits.}
  \label{fig:b-v-times-and-circuit}
\end{figure}

\subsection{Validation and evaluation of single-shot performance of stabilizer tableau simulation}

We cross-validate and compare the tableau's performance against Google Cirq \cite{kushnarev2025quditcirqlibraryextension}, a major software framework that supports qudit circuits through state vector simulation.  

\paragraph{Evaluation of correctness}
We empirically validated our simulator against Google Cirq's statevector simulation by asserting that they produce identical measurement outcome distributions on the same random Clifford circuits.
In particular, we ran the tableau and Cirq for 800 shots (sampled naively via repeated shots) over 1000 random Clifford circuits of a fixed prime dimension.  The depths ranged from 5 to 1000.  Let the distribution of measurement counts on any single circuit be $P$ and $Q$ for the tableau and Cirq, respectively.  
We verify that $\sum_{\ket{\psi}} P(\ket{\psi}) = \sum_{\ket{\psi}} Q(\ket{\psi}) = 1$.  Finally, we calculate the \textit{total variation distance} (TVD) between $P$ and $Q$, where 

\begin{equation} \label{TVD}
    \text{TVD}(P, Q) = \frac{1}{2} \sum_{\text{outcome } \ket{\psi}} \vert P(\ket{\psi}) - Q(\ket{\psi}) \vert
\end{equation}

We say distributions $P$ and $Q$ match when TVD$(P, Q) < 0.2$ on all of our random circuits.  The tableau passed this test.

\paragraph{Performance evaluation via random circuits}
A comparison between the tableau and Cirq of average runtime on random circuits is a straightforward task.  Figure \ref{fig:init-measurement} graphs the time to calculate single shot of a quantum circuit averaged over 10 random circuits of depth 1,000 swept over dimensions $d = 3, 5, 7$. There is a clear exponential dependence on the depth of the circuit in the Cirq simulation times.  
This dependence is exacerbated by the qudit dimension, indicative of state vector methods involving multiplying matrices of side length $d^n$.  On the other hand, tableau runtime is bounded by $O(n^2)$ with no apparent dependence on the dimension, reflecting the complexity of a single tableau operation depending only on its length. 

Nonetheless, the raw averages for Cirq runtimes contained outliers which likely depended on the size of the \textit{largest subset of qudits connected by CNOTs}.  
The complexity of a quantum measurement depends on the state space needed to describe it, and CNOTs turn single-qudit \textit{only} measurements (that live on a $d$-dimensional space) into multi-qudit measurements (that live on $d^2, d^3, \dots, d^n$ space).  
A proper comparison between Cirq and the tableau comes down to using a test family of test circuits that can precisely tune this measurement complexity, and the \textit{Bernstein-Vazirani} \cite{bernstein_vazirani} algorithm provides a such a circuit.

\paragraph{Measurement performance evaluation via Bernstein-Vazirani}
Bernstein-Vazirani (B-V) is a standard elementary quantum algorithm to solve the following problem: given a secret string (of $d$-valued dits) $s$ and a function $f(x) = x \cdot s$ (where $\cdot$ is the dot product), determine $s$ with minimal queries to $f$.  
The protocol to solve it is described by the circuit in Figure \ref{fig:b-v-circuit}.  
 The B-V oracle (gates that represent $f$ as an input) connects qudits to an ancilla if the qudit index correspond to a nonzero value in the bitstring (or ditstring).  One may control the measurement complexity simply by choosing the number of 1's in the secret string. 

Figure \ref{fig:b-v-times} depicts the performance of both simulators in evaluating 40-qudit B-V with random bitstrings of a fixed amount of 1s. Cirq's exponential dependence on the dimension of the largest entangled subspace completely eclipses the tableau runtime, which is only sensitive to the system size (and at most quadratic in it).

\paragraph{Local gate performance evaluation}  

\begin{figure}[t]
  \centering
  % center the main caption text
  \captionsetup{justification=centering}

  \begin{subfigure}{\columnwidth}
    \centering
    \includegraphics[width=0.9\linewidth]{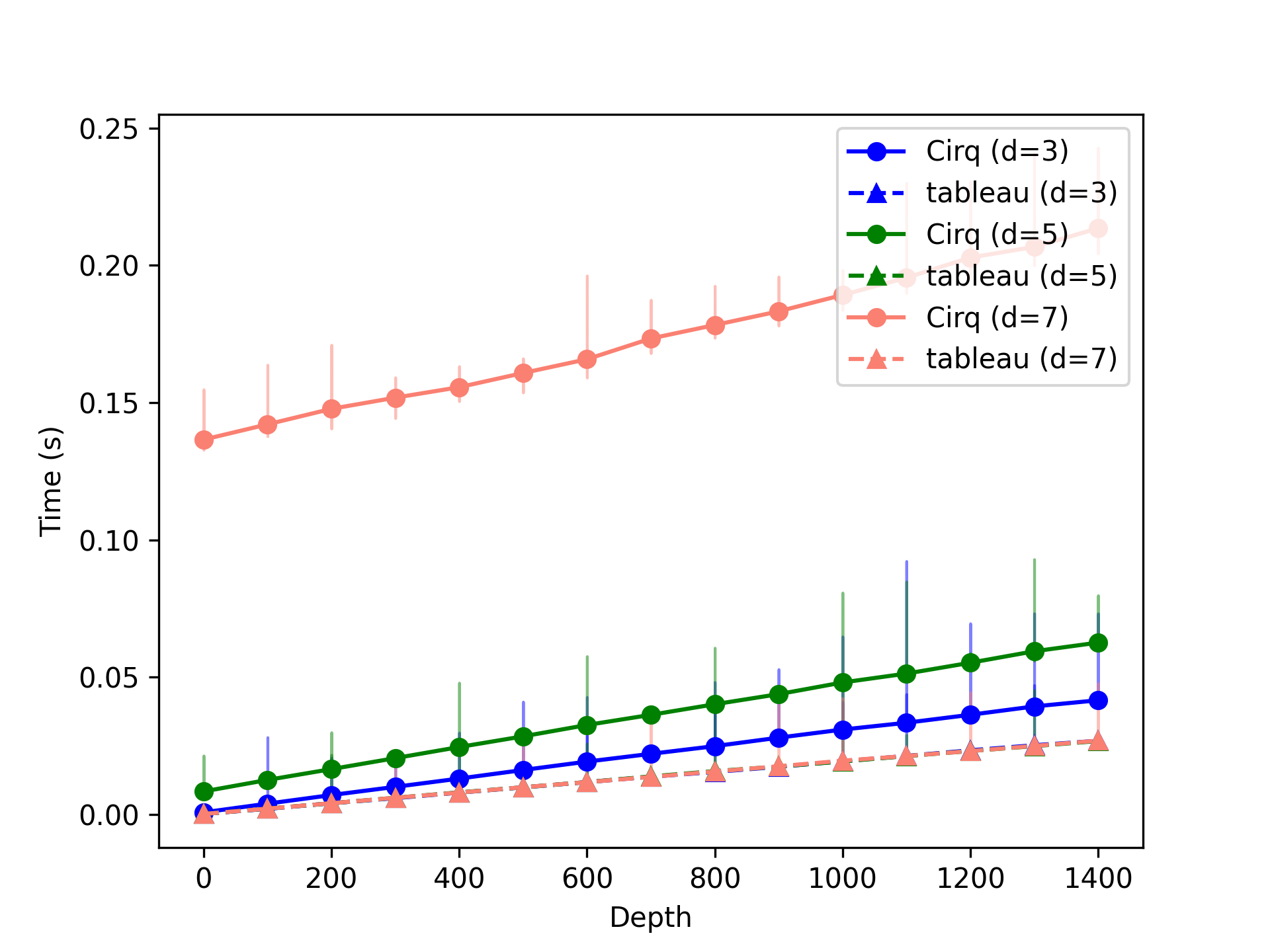}
    \caption{Average runtimes for Cirq and the tableau (over 100 trials at each depth $M$) of evaluating the circuit depicted below.}
    \label{fig:local-gate-test-results}
  \end{subfigure}

  \begin{subfigure}{\columnwidth}
    \centering
    \includegraphics[width=0.95\linewidth]{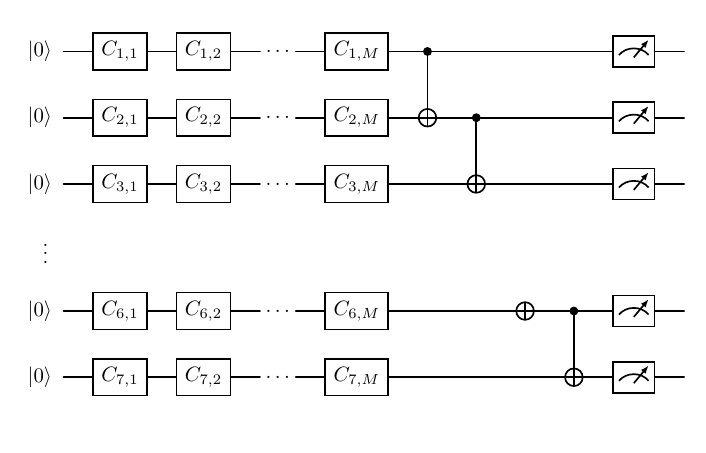}
    \caption{Local gate test over 7 qudits.  After each qudit undergoes $M$ random Clifford gates, every qudit is entangled via a chain of $CNOT$s before a round of measurement.}
    \label{fig:local-gate-test-circuit}
  \end{subfigure}

  \caption{Local gate test: runtime and circuit.}
  \label{fig:local-gate-test}
\end{figure}

Figure \ref{fig:local-gate-test} depicts a test over fixed $d = 7$ and a single measurement round of a fully entangled state which compares the simulators' abilities to apply gates on small systems.  The runtimes are all linear in the depth, but the tableau performance has no dependence on the dimension while Cirq's runtime increases greatly as the dimension increases.

% \begin{figure}[h]
%     \centering
%     \includegraphics[width=1\linewidth]{first_meas_eval_FINAL.png}
%     \caption{\textbf{Average runtime (over 100 trials at each depth $M$) of evaluating a 7-qudit circuit consisting of: $M$ local single qudit gates on each qudit, a chain of CNOTs to entangle the qudits, and finally a single terminal measurement. }}
%     \label{fig:local-gate-test}
% \end{figure}

\section{Evaluation and application of Pauli frames for qudit FTQC benchmarking}
% \yipeng{This is the second of two mini-papers in this manuscript.}

\begin{figure}[t]
    \centering
    \includegraphics[width=0.9\linewidth]{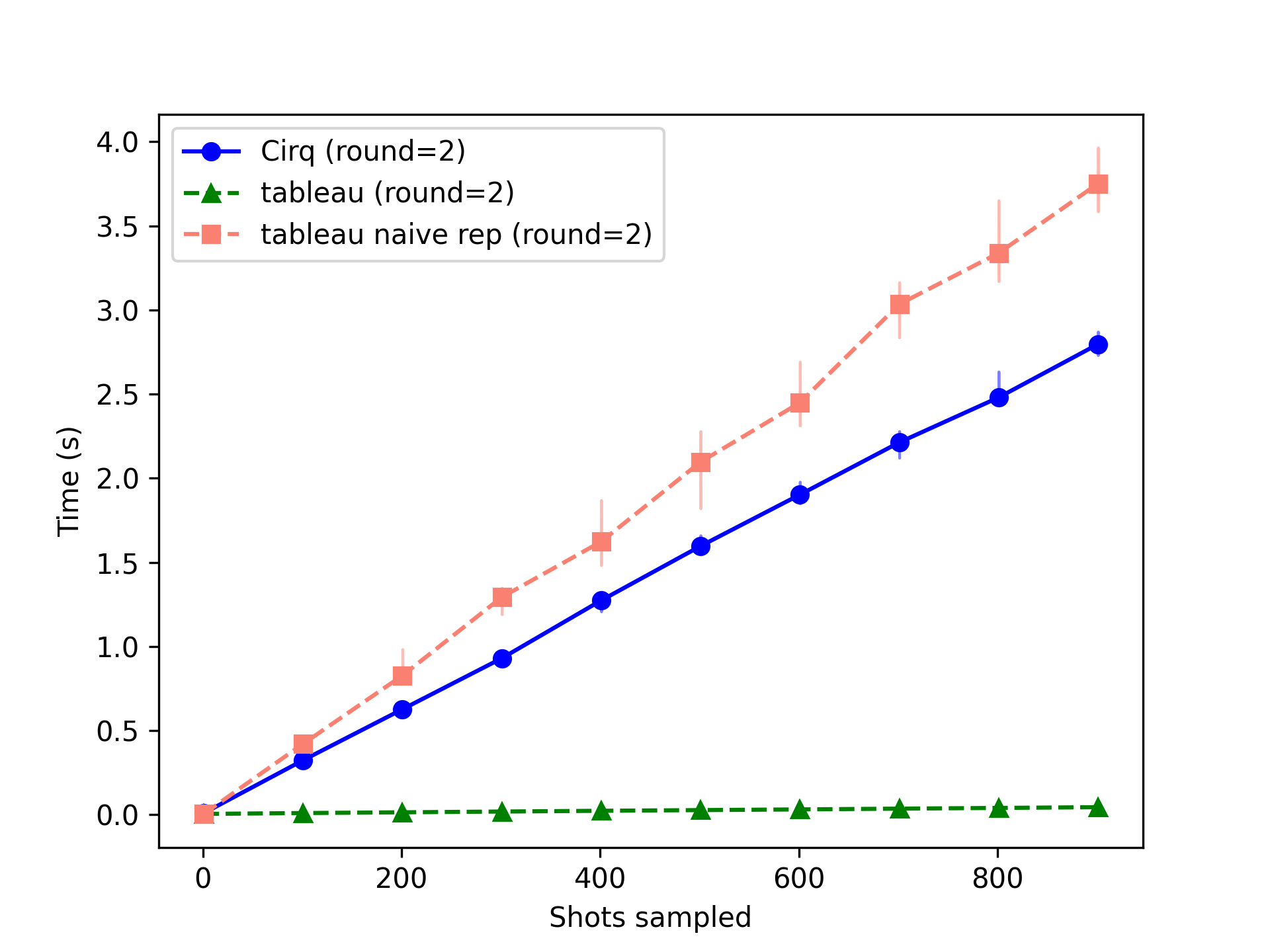}
    \caption{Time to sample the circuit from Figure \ref{fig:local-gate-test-circuit} using Cirq (blue, solid, circles), naive repeated runs with the tableau (orange, dotted, squares), and the Pauli frame tableau sampler (green, dotted, triangles).}
    \label{fig:sampling_method_comparison}
\end{figure}

Although stabilizer computation is faster than state vector simulation, naive repeated simulation of circuits with different noise trajectories is prohibitively expensive.  We use \textit{Pauli frame sampling} to compute many shots of a qudit circuit, which are then benchmarked against Cirq for speed.  This technique is the basis of efficient \textit{qubit} stabilizer circuit sampling in Stim \cite{Gidney_2021}.  Figure \ref{fig:sampling_method_comparison} depicts the need for effective sampling, as the Cirq sampler outperforms tableau repetition, and the tableau sampler completely outperforms the Cirq sampler.

\subsection{Implementation of qudit Pauli frames for Monte Carlo Pauli error channel simulation}

Pauli frames are Pauli operators that represent random rotations ~\cite{Knill_2005} that are aggregated together in block form.
We use frames to simulate the randomness from noise gates and from non-deterministic $Z$-measurements using only a single noiseless \textit{reference shot} from regular tableau simulation.
The key limitation of this method is that our noise consists strictly of probabilistic Pauli channels, which struggle to model other common types of noise noise like leakage errors  \cite{manabe2025efficientsimulationleakageerrors} despite their ubiquity in QECCs ~\cite{Approximation_of_realistic_errors,qec_threshold_approximate_errors}.

\subsubsection{Pauli error channels} As mentioned in section 3, the tableau supports single qudit \textit{flip, phase,} and \textit{depolarizing} noise channels given a parameter \verb|prob|.  All channels act on the tableau by $I$ with probability $(1 - \verb|prob|)$.  With probability \verb|prob|, flip and phase errors enact a random pure $X$-type and pure $Z$-type error, respectively, while depolarizing noise applies a random arbitrary (non-identity) Pauli gate.
\subsubsection{Pauli frame tableaus} 

The frame tableau is simply a table of stacked, \textit{phaseless} Pauli operators.

\[
    \left[ \begin{array}{ccc|ccc}
        x_{0, 0} & \dots & x_{0, n - 1} & z_{0, 0} & \dots & z_{0, n - 1} \\
        \vdots & \ddots & \vdots & \vdots & \ddots & \vdots \\
        x_{s, 1} & \dots & x_{s, n} & z_{s, 0} & \dots & z_{s, n - 1}
        \end{array} \right]
\]

The length $s + 1$ is the number of additional samples to generate from the single reference shot.  This reference obtained by tableau simulation that \textit{by ignores all noise gates}.
Each row in this frame tableau represents the random rotations of a distinct shot. The procedure for initializing and evaluating a single frame (row) is as follows:

\begin{enumerate}
    \item Initialize frame $i$ with the block form of $Z_1^{k_1} \otimes Z_2^{k_2} \dots \otimes Z_n^{k_n}$, where each $k_i \in \{ 0, \dots, d - 1 \}$ is random.

    \item If the frame encounters a \verb|RESET| on qudit $j$, repeat n step 1 \textit{only} for the $j^{\text{th}}$ entry of each frame.  Otherwise, conjugate the frame with the usual rules.
    
    \item If the frame encounters a noise $X^k Z^l$ on qudit $j$, multiply it into the frame.  This means $x_{i, j} \gets x_{i,j} + k \mod d$ and $z_{i, j} \gets z_{i,j} + l \mod d$.

    \item When measuring on qudit $j$, record the value as $r + x_{i, j} \; (\text{mod } d)$, where $r$ is the value of that measurement recorded during the reference shot.  Randomly re-initialize $z_{i, j}$.
\end{enumerate}

  The evolution of the shots above, which is just arithmetic over \textit{columns}, is effectively vectorized by the Numpy library \cite{harris2020array}.  The frame calculates \textit{deviations} off the reference shot whose statistics are governed by the randomness in the circuit. Let us examine a frame tableau for $H\ket{0}$. Suppose our reference shot measures $H\ket{0}$ as $k$ ($Z$-eigenstate $\ket{k})$.  We initialize our frame tableau with a random $Z$-block, which evolves according to our procedure:

  \begin{gather*}
    \left[ \begin{array}{c|c}
        0 & z_{00} \\
        0 & z_{10} \\
        \vdots & \vdots \\
        0 & z_{s0} \\
        \end{array} \right]
        \xrightarrow[\texttt{H 0}]{}
        \left[ \begin{array}{c|c}
        \textcolor{red}{-z_{00}} & 0 \\
        \textcolor{red}{-z_{10}} & 0 \\
        \vdots & \vdots \\
        \textcolor{red}{-z_{s1}} & 0 \\
        \end{array} \right]
        \xrightarrow[\texttt{M 0}]{}
        \left[ \begin{array}{c|c}
        -z_{00} & z'_{00} \\
        -z_{10} & z'_{10} \\
        \vdots & \vdots \\
        -z_{s0} & z'_{s0} \\
        \end{array} \right]
\end{gather*}

where the red values are all distinct offsets to add to our reference measurement of $k$.  We see the distribution $\{k - z_{j 0} (\text{mod } d)\}_{0 \leq j \leq s}$ is uniformly random as expected.  This example illustrates the role of \textit{random $Z$-block initializations} in steps 1, 2, and 5 above.  The procedure is $O(ns + gs)$, which bounds the time it takes to set up the frame tableau and to collect shots in a circuit of $g$ gates.

 %Implications for speed
\subsection{Validation and evaluation of multi-shot performance of qudit Pauli frame simulation}

We validate the correctness of sampling and error channels, and we benchmark performance of the sampler against Cirq.

% \subsubsection{Validation of Pauli frame simulation}
% All tests are available on the simulator github repository as \verb|pytest| modules.  

\paragraph{Validation of sampling without noise}
Given $n$ and any dimension $d$, we collect samples of random depth 1000 $n$-qudit circuit (with no noise gates) using naive repetition and the sampler.  We obtain empirical distributions $P$ and $Q$ over a sample size $s$, respectively, and we assert $\text{TVD}(P, Q) < 0.02$ as matching expected output.  The tableau passes this test.
  
\paragraph{Validation of error channels}
The validation procedure for error channels is very similar.  Given a sample size $s$, a length $M$, and any dimension $d$, we apply $M$ random Clifford gates to $\ket{0}$, apply the noise gate, and measure.  We apply the noise gate via the sampler as well as construct $s$ separate circuits of random Pauli gates that implement the effect of the respective channels directly.  These two methods yield distributions $P$ and $Q$, respectively, and we assert $\text{TVD}(P, Q) < 0.02$ as matching expected output.  The tableau passes this test.

% The procedure to test single channel error randomly samples Pauli $Q$ that is \textit{not} $X$ or $Z$.  We prepare the eigen

% results in a uniform measurement distribution in a random Pauli (non-eigen) basis: \selfnote{revise for clarity and brevity.}

% \begin{enumerate}
%     \item Select a random error probability $p$, dimension $d$, a Pauli $Q = UZU^\dagger$ where $U\neq I, H$, and a sample size $s$.
    
%     \item Begin with an empty circuit of a single qudit and apply $U^\dagger$.  Apply the error channel, and measure in $Q$.  Sampling $s$ shots yields a probability distribution of outcomes $P$.

%     \item For phase and flip channels, let $Q$ be the distribution that assigns probability $1-p$ to 0 and $p / (d^2 - 1)$ to everything else.  For depolarizing, assign $(1 - p) + (d - 1)(p / (d^2 - 1))$ to 0 and probability $d\cdot  (p / (d^2 - 1))$ to everything else.  Assert that $\text{TVD}(P, Q) < 0.02$.
% \end{enumerate}

\begin{figure}[t]
    \centering
    \includegraphics[width=0.9\linewidth]{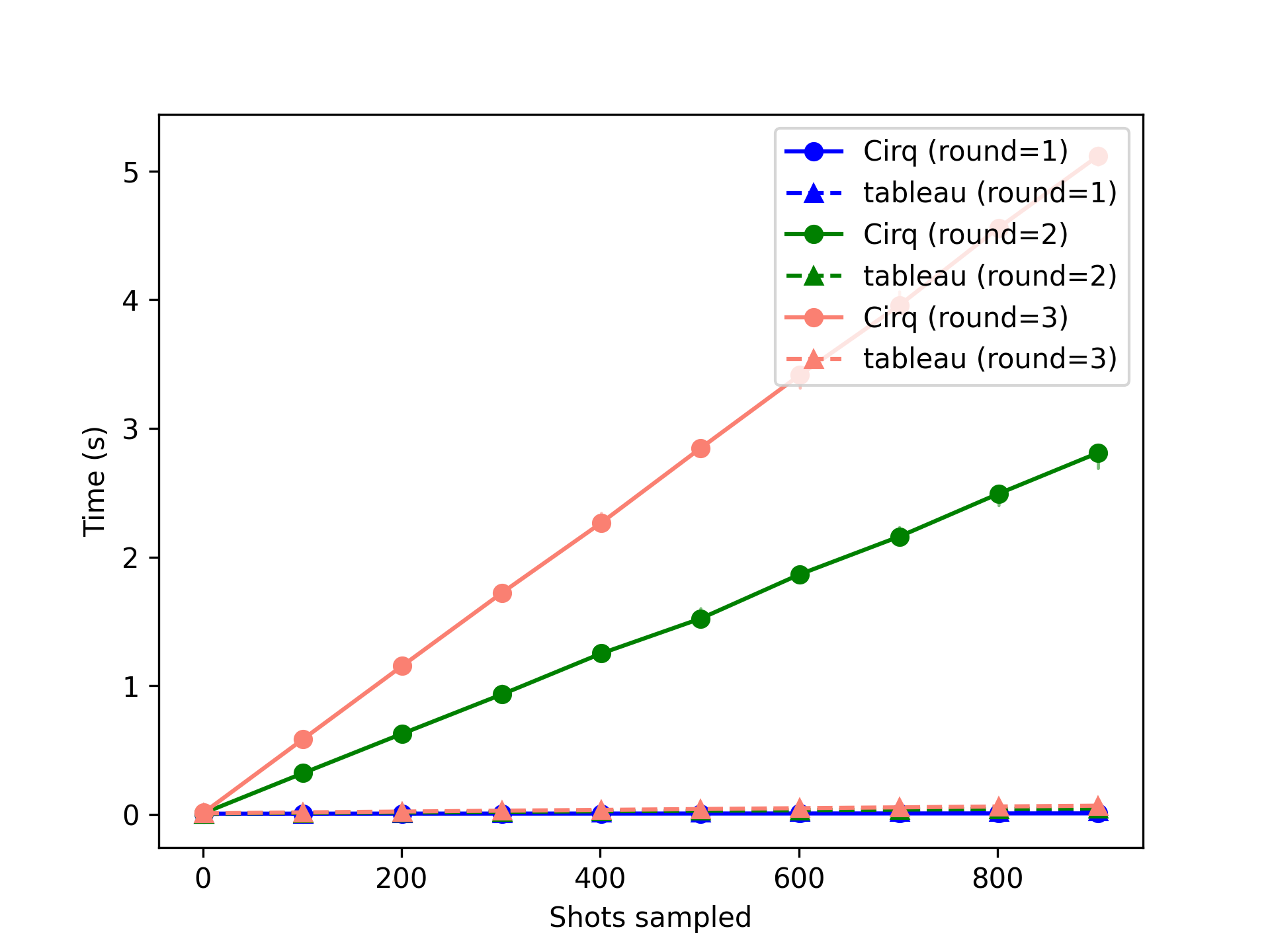}
    \caption{Above is the time it took Cirq and the tableau to sample shots of Figure \ref{fig:local-gate-test-circuit} as well that circuit concatenated with itself various times.  The number of times here are called \textit{rounds}.
    The runtimes per shot count have been averaged over 10 runs of sampling.}
    \label{fig:frametimes}
\end{figure}

\subsubsection{Evaluation of Pauli frame performance} 

In order to compare the sampling speeds between the tableau and Cirq, we run the two simulators on $n$-fold self-concatenations of the circuit in Figure \ref{fig:local-gate-test-circuit}.  Each fold introduces a new round of measuring an entangled 7-qudit state, which helps us compare sampling complexity between simulators while also comparing the overall contribution of sampling compared to collecting the reference measurement.

In 1 round (no self-concatenation), Cirq actually outperforms the tableau at collecting shots, suggesting that the statevector sampling \textit{alone} is comparable or better than frame sampling.  However, the measurement complexity still completely overtakes the sampling complexity seen in deeper rounds of self-concatenation.  There are small, unentangled systems where entangled statevector computation is fast to compute, but tableau simulation is \textit{always} cheap and certainly many orders of magnitude cheaper for large, entangled systems.

\section{Case study:  Logical error benchmarking and calculating the threshold of a five-qudit code}

The qudit tableau simulator was built ultimately as a tool with which to benchmark the error-correcting capabilities of qudit code families.  
An error correction protocol measures the code stabilizers with a heavily-structured, often repeated procedure, after which the resulting data is passed to a \textit{decoder}, which outputs error (or equivalently, recovery) information about the program.
After repeated trials with varying parameters, the result of the protocol is a curve relating the physical error rate(s) associated with each  or idle moment to the logical error rate of the system (i.e. the rate at which the corrected program output is indeed error-free).
Finally, by self-concatenating the code multiple times and performing the benchmarking protocol, the approximate intersection of the curves, called the \textit{threshold}, is the physical error rate at which the code can correct all errors in our assumed error model \cite{doi:10.1137/S0097539799359385}.

Recent work on simulating qudit codes has yielded a threshold for a $[[5,  1,  3]]_q$ code, which is a $q$-dimensional generalization of the qubit perfect code \cite{keppens2025quditvsqubitsimulated}, but the results, which were obtained using Cirq, quickly became infeasible to run as the dimension increased (namely $q = 7$). 
In order to demonstrate the simulator and sampler's utility, we will extend the paper's analysis of belief matching (BM) versus regular minimum-weight perfect matching (MWPM) decoding \cite{Higgott_2023} to the $q = 7$ variant of the code, and we will calculate a threshold for the code under \textit{circuit-level noise} using BM.

\subsection{Evaluation of logical error rate}

A simple logical error rate (LER) $P_L$ is a curve obtained by the following procedure: initialize the state as some $\ket{\Psi}$, apply depolarizing noise on each physical data qudit with a probability of physical error $p_c$, and then measure stabilizers.  
The error-and-measurement procedure may be repeated in the circuit over many rounds to test the code's robustness against noise buildup over time.
We end the test by measuring logical operator $L$ such that initial state $\ket{\Psi}$ is an $l$-eigenstate of $L$.  
Each error configuration in the circuit shifts the logical measurement outcome $l$ by an integer, meaning there is a one-to-one map between errors and $L$-measurements.  
Our correction protocol is performed ``offline,'' where we collect samples of stabilizer measurements and of $L$, feed the stabilizer measurements into a decoder, and map their error output to a shift in $L$.  If the decoded shift matches the shift observed in the sampled shot, then the sample is marked ``correct'' and ``incorrect'' otherwise.  The work to feed qudit data into qubit decoders like MWPM \cite{higgott2021pymatchingpythonpackagedecoding} and BM \cite{Higgott_2023} to successfully correct qudit errors (by introducing new nodes for the non-binary syndrome flips) was established in \cite{keppens2025quditvsqubitsimulated} and reproduced here exactly.

\begin{figure}[t]
    \centering
    \includegraphics[width=0.9\linewidth]{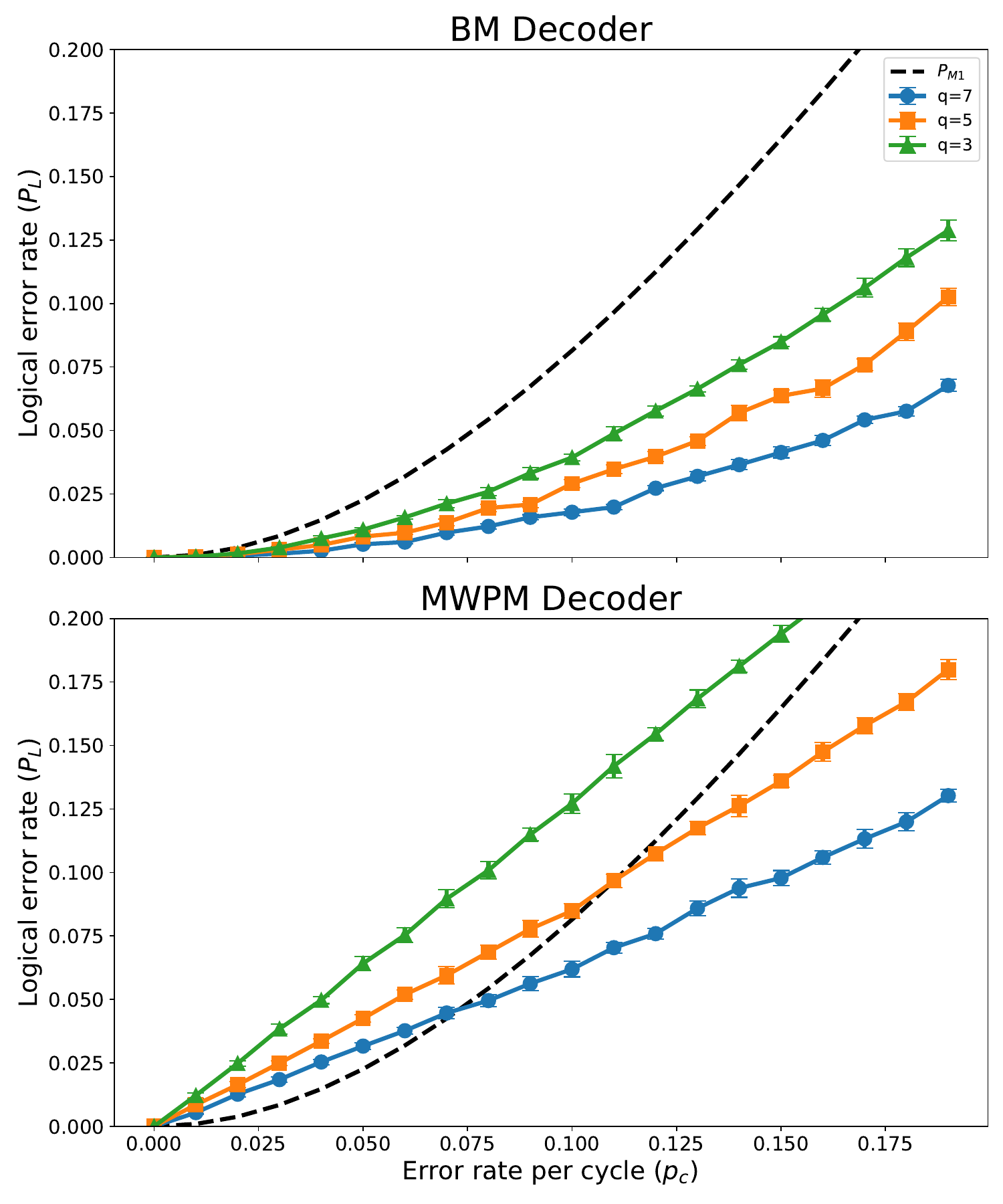}
    \caption{Logical error rate (LER) graph for MWPM and BM decoders on the $[[5, 1,  3]]_q$ code for $q = 3, 5, 7$.  Each data point is averaged over 100,000 shots.  The points on which the curve lies under the under the physical error rate (black dotted line) represent effective error correction.}
    \label{fig:noise-cap}
\end{figure}

The $[[5,  1,  3]]_q$ code has the following stabilizers:

\begin{equation} \label{codespace-stabs}
    \begin{gathered}
    G_1 = I \otimes Z \otimes X  \otimes X^\dagger \otimes Z^\dagger   \\
    G_2 = Z^\dagger \otimes I \otimes Z \otimes X  \otimes X^\dagger  \\ 
    G_3 = X^\dagger \otimes  Z^\dagger \otimes I \otimes Z \otimes X  \\
    G_4 = X \otimes X^\dagger \otimes  Z^\dagger \otimes I \otimes Z \\
\end{gathered} 
\end{equation}

Our LER benchmark measures logical transversal operator $\overline{Z} = Z_0  Z_1  Z_2  Z_3  Z_4$ onto an ancilla.

% \selfnote{Discuss results in brief explaining the match in shape but the difference in crossover in analysis of error binning (logical shifts) versus no hook errors.}

Figure \ref{fig:noise-cap} displays the LER curves in $q = 3, 5, 7$ computed with Sdim.  The curves for $q = 3, 5$ match the shape of the original experiment, and the tableau was able to collect a sufficient number of samples to plot the $q = 7$ in significantly less time than statevector simulation.  Additionally, the tableau simulation bins successes and failures based on shifts to a logical operator, meaning that physical errors that commute with the $\overline{Z}$ have no effect on the final measurement in contrast with the statevector overlap evaluation in the original experiment.  This logical operator approach is simultaneously more representative of real quantum computer output and yields a slightly more optimistic logical error rate than the original result.  As a final note, we observe here that contrary to the original paper's claims, both MWPM and BM decoders report strictly lower logical error curves as $q$ increases, suggesting that a more careful analysis into edge weights is necessary to truly ascertain the potential advantages or disadvantages of decoding qudit codes.  The ability to rapidly produce these simulations is vital in settling the matter.

\begin{figure}[t]
    \centering
    \includegraphics[width=\linewidth]{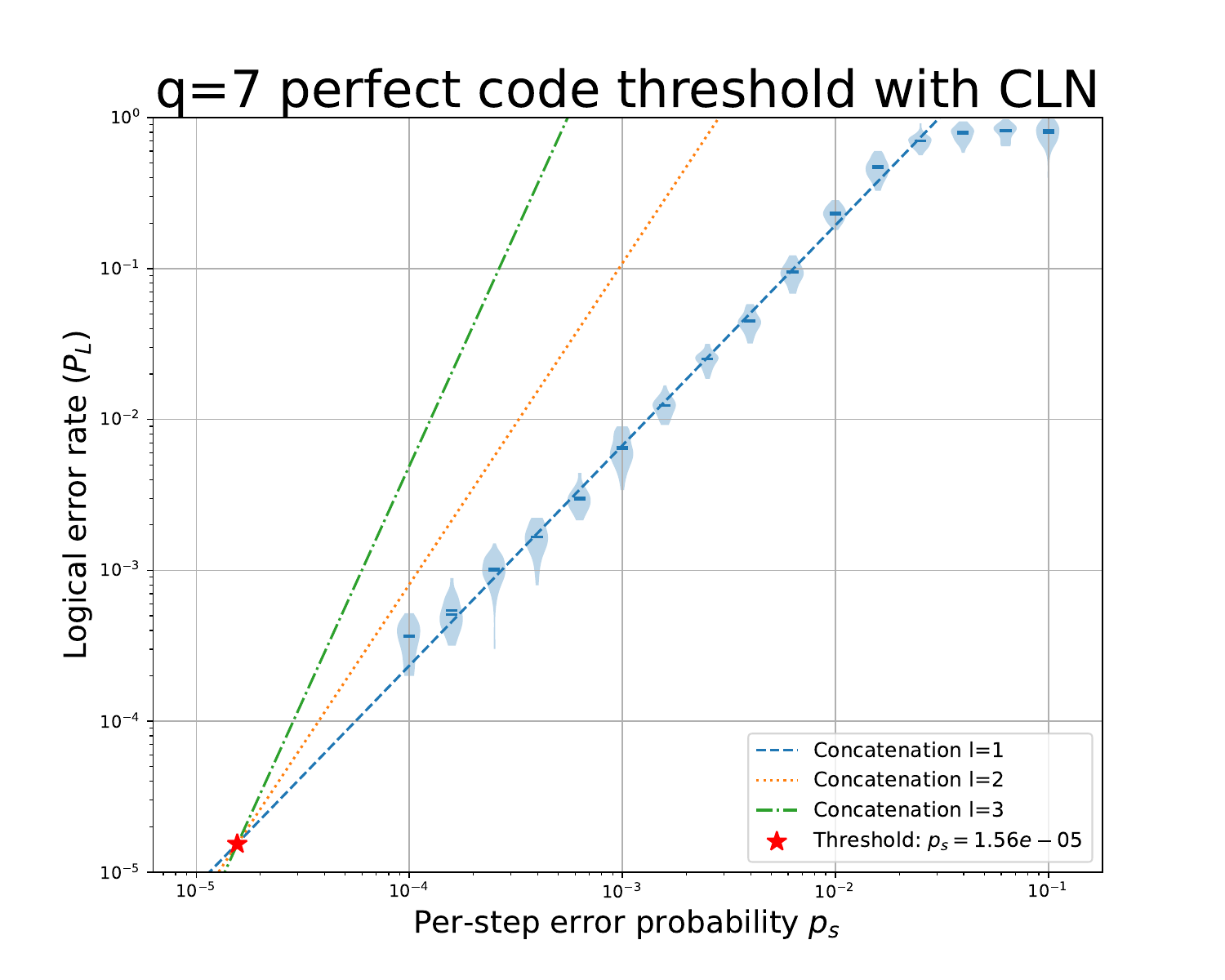}
    \caption{Threshold estimation using curve fit concatenations of the $[[5, 1,  3]]_7$ code under circuit-level noise (idling, depolarizing gate errors, and measurement bitflips).  Each data point is averaged over 1.5 million shots and decoded using BM.}
    \label{fig:threshold-graph}
\end{figure}

\subsection{Evaluation of threshold with circuit-level noise}

To obtain a threshold, we perform an LER with three stabilizer measurement rounds prior to our terminal $\overline{Z}$ measurement.  State preparation is noiseless, but in addition to the usual LER depolarizing events, our first two rounds of syndrome extraction injects noise on the data qudits during both idling and gate operations, noise on ancilla qudits when gates are applied and a bit-flip noise channel prior to the $\overline{Z}$ measurement.  All channel probabilities are identically parameterized by a single physical error rate $p_s$.  

We stand by the original paper's justification to approximate the concatenated code error rates using a power law curve fit.  
The new logical operators under concatenation, loss of transversality, and potential decoding ambiguities necessitate attention that reaches beyond the scope of this demonstration.  
Our focus here is on the simulator's ability to output and process large ensembles of sampled shots.
Given our initial logical error rate $P^1_L(p_s) = ap_s^b$, where $a$ and $b$ are fit constants, we estimate the logical error rate of the $k^\text{th}$-level of self-concatenation by the recursive expression $P^k_L = P^1_L(p_s) \; \circ \; P^{k - 1}_L$.  The threshold is present in Figure \ref{fig:threshold-graph}.  At of the time of writing, this is seemingly the first threshold graph of a high-dimensional ($q>5$) code under circuit-level noise.
Finally, the timings of sampling shots of the (noisy) quantum circuits for these simulations can be seen in Table \ref{tab: timings}. As expected, sdim is several orders of magnitude faster in sampling than Cirq, making decoding the time-limiting bottleneck.\\

\begin{table}[h]
    \centering
    \caption{Times required to sample syndrome measurements with the noisy $q=7$ five-qudit codes in Sdim and in Cirq.}
    \begin{tabular}{c|cc} 
 & Sdim & Cirq \\
\hline
LER sample & $9.9 \;\cdot 10^{-5}$ s & $66$ s \\
Threshold sample & $2.3 \;\cdot 10^{-4}$ s & -- 
\end{tabular}\\
 \label{tab: timings}
\end{table}

\section{Case study: Validation of logical error benchmarking of a five-qutrit surface code}

The simulator's sampler was developed to support characterizing emerging qutrit hardware.  
We are motivated by the following: can we still meaningfully extract early fault-tolerant behavior from a noisy system using a code that \textit{detects} errors rather than \textit{corrects} them?  Such is currently more feasible for systems with a limited number of nearest-neighbor interacting qudits.  To that end, we prototyped a \textit{detection-only, postselecting} variant of \textit{logical randomized benchmarking} (LRB) \cite{combes2017logicalrandomizedbenchmarking} called LRB-D.  Initially, our study began with simulating the fidelity of a 5-qutrit surface code protected by a \textit{single round} of stabilizer measurement postselection on a 5-qutrit surface code.  We expect this protocol to actually decay \textit{faster} than a single physical qutrit as a 5-qutrit system is vulnerable to more errors than a 1-qutrit system, and the single postselection round would be inadequate to pick out non-erroneous shots of the circuit.  Nonetheless, this was our first step before developing the protocol past the scope of showcasing the simulator. We were able to only partially verify our predictions in Cirq, which took prohibitively long to run a \textit{simpler version} of the simulation.  On the other hand, the tableau was able to quickly reproduce the partial simulation as well as efficiently complete the full protocol. \\
\subsubsection*{Randomized benchmarking}

\begin{figure}[h]
    \centering
    \includegraphics[width=0.95\linewidth]{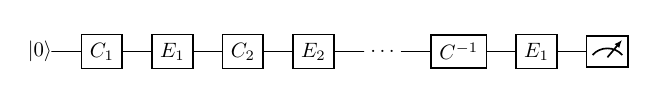}
    \caption{A typical RB circuit where $C_i$ are the sampled Clifford gates, $C^{-1}$ is the product of inverses of the preceding Clifford gate sequence, and $E_i$ are the errors after each gate.  Without the error operators, this circuit evaluates to identity on the $\ket{0}$ state.}
    \label{fig:RB}
\end{figure}

This decay parameter $\alpha$ may be obtained either from hardware experiments and from idealized simulations.  The work in ~\cite{Dankert_2009, Magesan_2011, Magesan_2012, Wallman2018randomized} establishes that a simulated $\alpha$ with access only to Clifford circuits and depolarizing noise channels can reasonably model noise processes on real hardware.  
\subsubsection*{Detection codes}

\begin{figure}[!h]
  \centering
  % center the main caption text
  \captionsetup{justification=centering}

  \begin{subfigure}{\columnwidth}
    \centering
    \includegraphics[width=0.88\linewidth]{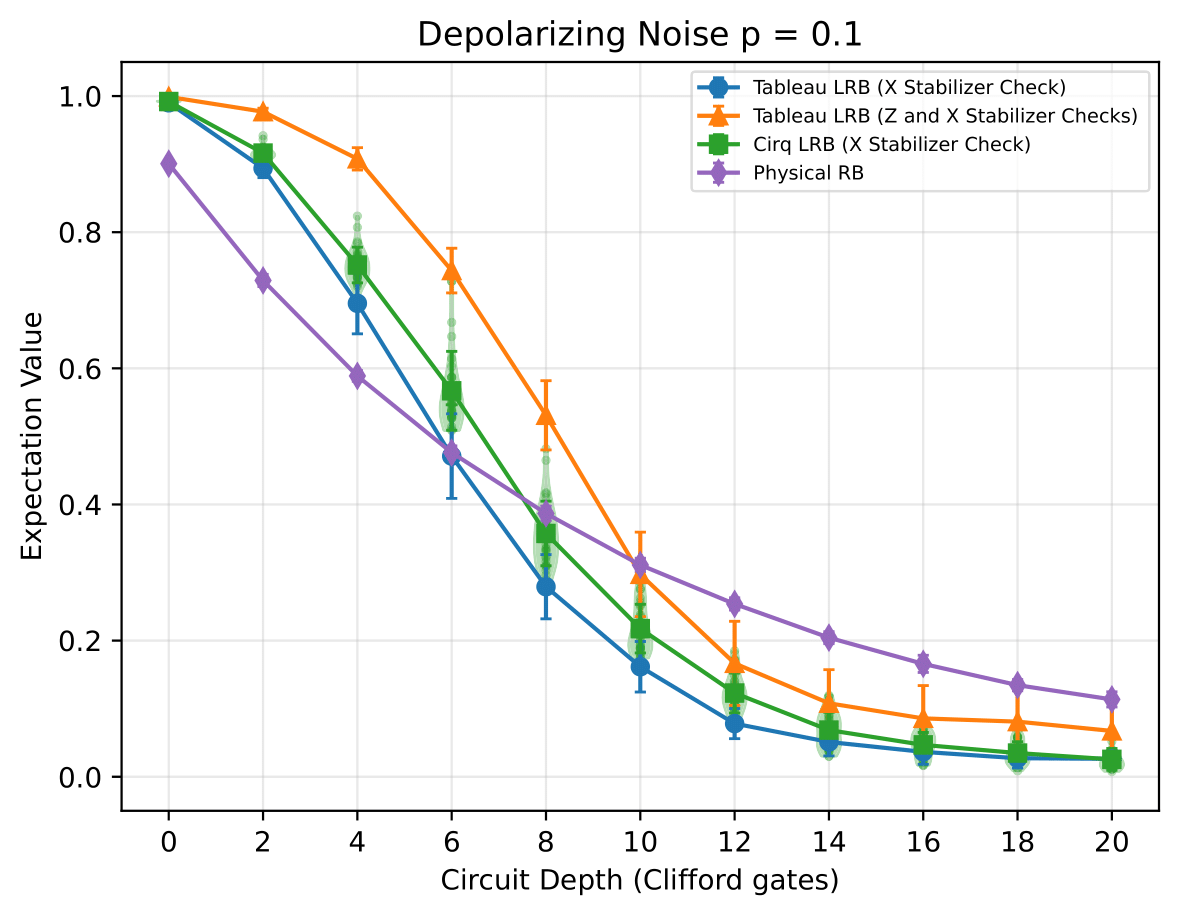}
    \caption{Tableau and Cirq fidelity curves of the LRB-D (detailed in Figure \ref{fig:LRB_circuit}) at a physical gate error rate of $p = 0.1$.  We were also able to extend the original experiment and simulate measuring all codespace stabilizers with minimal impact on performance.  The purple curve is an RB curve for a physical single qutrit.}
    \label{fig:LRB results}
  \end{subfigure}

  \begin{subfigure}{\columnwidth}
    \centering
    \includegraphics[width=0.85\linewidth]{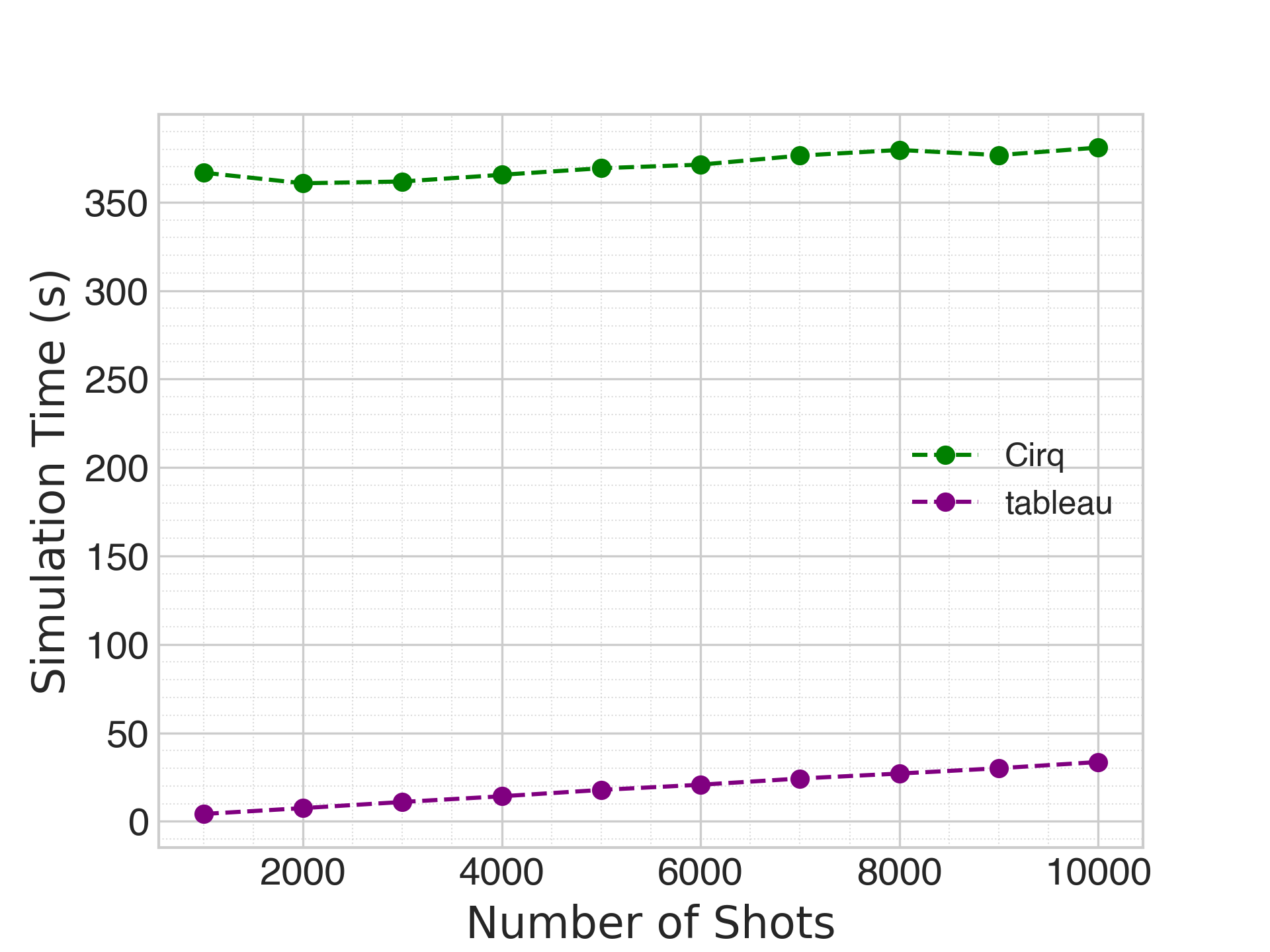}
    \caption{Simulation time per run of the 5-qutrit surface code LRB-D (at a fixed physical error rate of $0.1$, logical depths $0, 5, \dots, 20$, averaged over $30$ random Clifford circuits at each depth) versus sample size.  Tableau methods offer significantly improved runtimes even at modest system sizes.}
    \label{fig:sdimvcirqlrb}
  \end{subfigure}

  \caption{LRB-D fidelity curves and runtimes between Cirq and the tableau.}
  \label{fig:LRB-sdim-v-cirq}
\end{figure}

\begin{figure*}[h]
    \begin{subfigure}{.45\textwidth}
      \centering
      \includegraphics[width=0.95\linewidth]{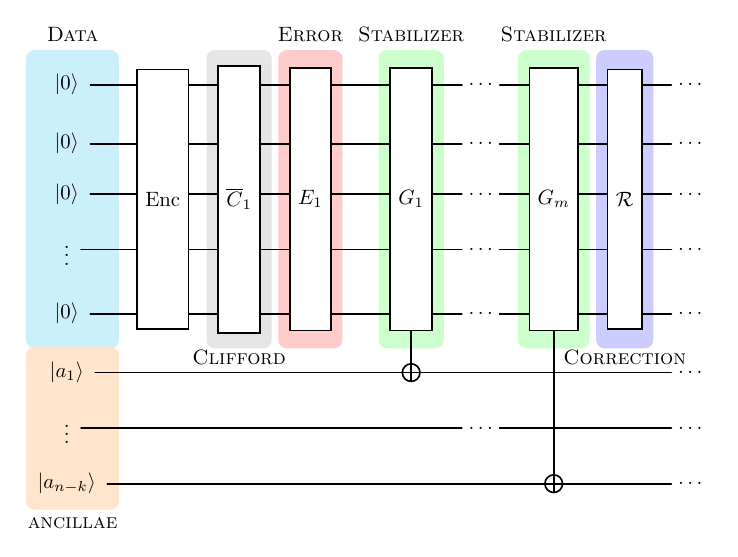}
      \caption{A single round of a logical randomized benchmarking circuit with syndrome extraction.  The state is prepared into an initial eigenstate of a logical operator $\overline{Q}$. After noisily inverting the circuit, we measure $\overline{P}$.}
      \label{fig:LRB round}
    \end{subfigure} \hspace{.03\textwidth}
    \begin{subfigure}{.5\textwidth}
      \centering
      \includegraphics[width=\linewidth]{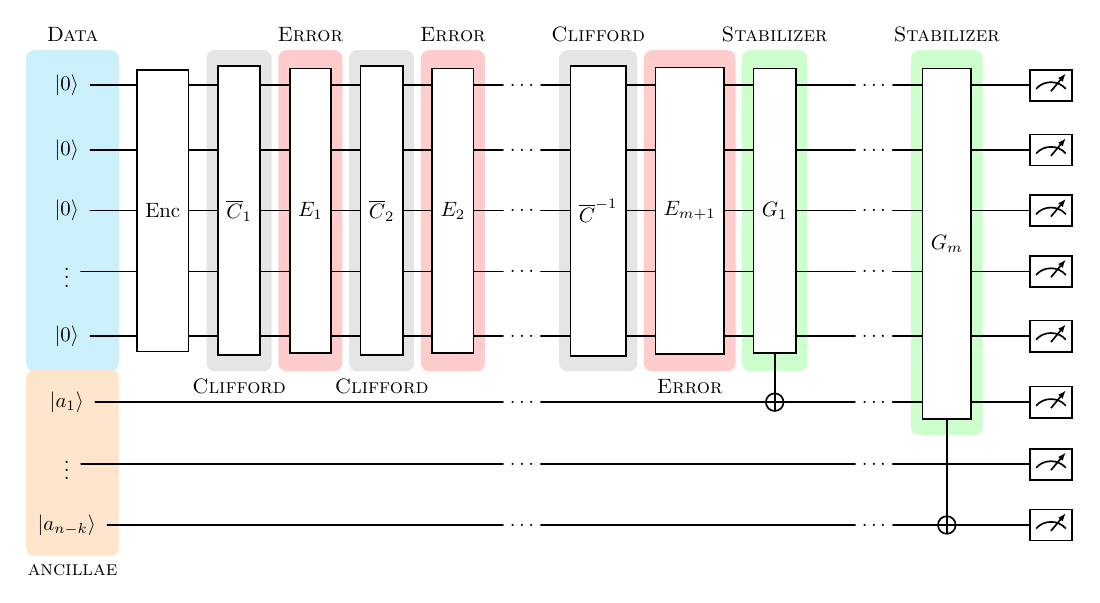}
      \caption{LRB with detection only (LRB-D).  There is only one round of terminal syndrome extraction used to postselect shots with no detected noise. The terminal $\overline{Q}$ measurement can be expressed as a sum of measurements on all data qudits.}
      \label{fig:LRB-D-protocol}
    \end{subfigure}
    \caption{Logical randomized benchmarking with non-destructive stabilizer
measurements.}
\end{figure*}

The \textit{Randomized benchmarking} (RB)~\cite{Knill_2008_randomized} procedure is a Monte Carlo method that, given error rates of a noisy quantum gate set, produces an $\alpha \in [0, 1]$ such that $f(D) = B \alpha^D$ approximates the \textit{fidelity}, or the probability that a circuit with noisy gates returns correct measurements, where $B$ is a constant and $D$ is circuit depth.  For simplicity, we assume all gates have the same physical error rate.

\begin{enumerate}
    \item Generate random circuits of the type in Figure \ref{fig:RB}, parametrized by depth and the number of random circuits at each depth.

    \item Simulate each circuit over a large constant number of shots.  For the $k^{\text{th}}$ random circuit of depth $D$, the (normalized) outcomes form a probability distribution $P_{D, k}$ over $\{ 0, 1, \dots, d - 1\}$.

    \item These probabilities yield \textit{fidelities} $f_{D_,k} = \left\vert \sum_{j = 0}^d P_{D, k}(j) \omega^j \right\vert.$

    \item Average the fidelities over $k$.  The resulting $f_D$ is the \textit{average fidelity of a depth } $D$ \textit{circuit} with this noisy gate set and takes the shappe of an exponential decay curve \cite{Wallman2018randomized}, and so a curve fitter can extract $\alpha$. 
\end{enumerate}

Detection codes are encodings where the stabilizer measurements can determine the presence of error but do not necessarily determine the error.  The work here considers a $[[5, 1, 2]]_3$ detection surface code \cite{Moussa_2016} defined by the following stabilizers:
\vspace{-6pt}

\begin{equation} \label{codespace-stabs}
    \begin{gathered}
    G_1 = Z_1 \otimes Z_2^{\dagger
    } \otimes Z_4^{\dagger} \;\;\;\;\;\;  G_2 = Z_2 \otimes Z_3 \otimes Z_5^{\dagger}\\ 
    G_3 = X_1 \otimes X_2 \otimes X_3^{\dagger} \;\;\;\;\;\; G_4 = X_2^{\dagger} \otimes X_3 \otimes X_5^{\dagger}\\
\end{gathered} 
\end{equation}

\subsection{Detection-only logical randomized benchmarking}

The \textit{logical randomized benchmarking} (LRB) protocol proposed by Combes et al.~\cite{combes2017logicalrandomizedbenchmarking} extends the RB framework to characterize the \textit{logical error rate} of an $n$-qudit code-protected system. In particular, the test circuits take stabilizer measurements after every logical Clifford operator and applies a correction.  Figure \ref{fig:LRB round} illustrates a single step of applying a single logical Clifford and error (depolarizing noise on all physical qudits affected by the Clifford parameterized by the physical error rate of each gate).  This is repeated for every logical Clifford The terminal measurement is replaced by a logical Pauli $\overline{Q}$ measurement for which the initial encoded state is an eigenstate.  From here, we process the measurements identically as in RB to obtain a decay curve for the logical fidelity.\\

However, full LRB is a demanding target for early qudit hardware. It requires syndrome extraction after every logical Clifford, a correction rule for each observed syndrome, and physical error rates low enough that repeated correction improves the logical channel. The decoding problem is especially delicate in the small, distance-two systems considered here: the syndrome can reveal that an error has occurred, but it generally does not provide enough information to identify a unique correction. In addition, a practical qudit decoder must distinguish between different nontrivial Pauli powers, incorporate the hardware-specific noise model, and apply corrections in real time without introducing additional logical faults. For these reasons, it is more realistic at this stage to ask what can be learned from error detection and postselection alone. We therefore consider a detection \textit{only} variant, called \textit{LRB-D}, detailed in Figure \ref{fig:LRB-D-protocol}.  The key modifications are that \textit{we only perform a single terminal syndrome extraction} prior to measuring $\overline{Q}$, and \textit{we postselect measurements with the all zeroes syndrome}, which certifies the code detected no error.  These postselected measurements are then processed identically as in RB.  This protocol is geared towards early FTQC efforts on qudit hardware.
% \begin{figure}
%     \centering
%     \includegraphics[width=0.88\linewidth]{LRB_sdim_vs_cirq-anon.png}
%     \caption{Sdim validation of the Cirq implementation of LRB-D (detailed in Figure \ref{fig:LRB circuit}).  We were also able to extend the original experiment and simulate measuring all codespace stabilizers with minimal impact on performance.}
%     \label{fig:LRB results}
% \end{figure}

% \begin{figure}
%     \centering
%     \includegraphics[width=0.85\linewidth]{figures/sdim_v_cirq_LRB.png}
%     \caption{Simulation time per run of LRB-D (at a fixed gate-independent noise parameter, over $30$ random Clifford circuits and their subcircuits) on the folded detection code versus the number of samples.  Even on a 5-qutrit state, tableau methods offer significantly improved runtimes.}
%     \label{fig:sdimvcirqlrb}
% \end{figure}

\subsection{Evaluation of the noise characterization}  
Initial efforts to study LRB-D involved a sweep of 90 error probabilities.  Per probability, we sampled $30$ Clifford circuits at depths $[0, 4, 8, 12, 16, 20]$ at $10,000$ shots per circuit.  The high shot count was necessary to produce a statistically significant amount of postselected shots.  Measuring all the codespace stabilizers was prohibitively expensive in the Cirq simulation, so we restricted the postselection criteria to only syndromes from stabilizers $G_3, G_4$.  The tableau was able to do this as well as a simulation measuring all stabilizers. \\

Both simulators' LRB-D curves match (within their error bars), seen in Figure \ref{fig:LRB results}.  The full-stabilizer tableau protocol yielded higher fidelities, seen in orange.  Nonetheless, all the LRB-D curves decay much faster than a single physical, unprotected qutrit (in purple), as expected.  The single round of postselection cannot adequately protect the logical state once physical errors build into logical errors during increasingly deeper error trajectories prior to postselection.

\section{Conclusion and research directions}

We have built a quantum circuit simulator capable of evaluating high-dimensional stabilizer circuits tailored towards error correction research.
We provided the theoretical basis behind its construction as well as proof of correctness for our deterministic measurement, which builds off prior work in stabilizer simulation  \cite{Aaronson_2004, gottesmanhighdim}.
The gate behavior are exhaustively validated against statevector simulation for correctness and benchmarked for speed.
The asymptotic advantage of tableau simulation against statevector simulation of a single shot (in prime dimensions) is visible in practice even for a modest number of qudits.  

Additionally, we built an efficient sampler for Pauli error channels, and this sampler is at least as fast or (often) much faster than both naively repeating shots or sampling from statevector simulation.  The tableau's ability to efficiently sample large ensembles of entangled stabilizer states makes it invaluable in the essential workflow of high-dimensional FTQC research.  

We used the simulator to extend a recent benchmark of high-dimensional error correction beyond the prior limits of tractability in a $q = 7$ system with fewer than 10 qudits.  
The more optimistic logical error rates, partly due to the use of the preferred logical-operator criterion, appear to challenge the conclusions presented in \cite{keppens2025quditvsqubitsimulated}, suggesting the need for further work and more careful analysis of decoder weights.
Furthermore, we used the tableau to extract a threshold for a $q = 7$ code under \textit{circuit-level noise}, which is a task that has never been performed in the literature to date.  This demonstration highlights a new novel task that we have aimed to make as mundane as is required for serious engineering of qudit systems. 

Finally, we used the simulator to validate the initial stages of developing a new logical benchmarking protocol using postselection on error-detecting codes. Not only was our simulator able to reproduce the results for a limited
statevector simulation, it was hundreds of times faster and was able to slightly extend the experiment.  This LRB-D study should therefore be viewed as a validation and feasibility experiment rather than a complete characterization
of the protocol.  In particular, terminal detection can produce decay curves with non-Markovian structure that are not always well summarized by a single exponential fit.  A dedicated future LRB-D study will address this limitation by
examining interval-check postselection, where stabilizer measurements are performed after fixed blocks of logical Clifford gates rather than only at the end of the sequence.  This intermediate form of detection may suppress some of
the accumulated memory effects seen in terminal-only postselection while still avoiding the full overhead of real-time correction. Future work will compare different detection cadences, quantify the resulting postselection bias, test more flexible fitting models, and include more realistic hardware noise in order to determine when the extracted logical decay parameter is a reliable early-FTQC benchmark.

In the future, we would want to extend the simulator past the bare minimum needed for QECC research.  The error map in our case study is a limited functionality proxy for a full detector-error model, with which we would be able to perform complex, sophisticated, and experimentally relevant simulations of fault-tolerant protocols with novel codes and decoders.  These tools would be would be in service of interrogating high-dimensional decoding in a more precise and exhaustive manner.

% \paragraph{FTQC primitives and a maturing API}
% As stated in the beginning, this simulator is \textit{only} the bare minimum.  There is a long line of work yet to be realized on the way to supporting qudit FTQC, such as developing standard workflows, pipelines, and libraries.

% \cite{Simulation_Dominated_Clifford}
% Also desirable is an ability to simulate complicated noise that significantly affect accurate physical error thresholds such as spatially correlated noise~\cite{Sarovar2020detectingcrosstalk,Huang2020AlibabaCQ,PhysRevLett.126.230502} and temporally correlated non-Markovian noise~\cite{Gulacsi2023SignaturesON}.

\newpage

% use the ACM bibliography style
\bibliographystyle{IEEEtran}
\bibliography{references,stabilizers,qudits,qecc,noise,software}
%\input{main.bbl}

% \begin{appendices}
% \input{A_1_Qudit_Deutsch_Cartoon}
% \input{A_2_Prime_Measurement_is_Gaussian_Elim}
% \input{A_3_Non-Prime_Tableau}
%\input{A_4_Non-Prime_Measurement_is_SNF}
% \end{appendices}

\end{document}